\documentclass{jfm}
\usepackage{graphicx}
\usepackage{epstopdf, epsfig}
\usepackage[utf8]{inputenc}
\usepackage[english]{babel}
\usepackage[T1]{fontenc}
\usepackage{lmodern}
\usepackage[dvipsnames]{xcolor}
\usepackage[colorlinks=true,linkcolor=blue,citecolor=blue,filecolor=blue,urlcolor=blue]{hyperref}

\usepackage{amssymb}
\usepackage{bm}
\usepackage{lastpage} 
\usepackage{setspace}
\usepackage{fancyhdr} 
\usepackage{graphicx}
\usepackage{tikz}
\usetikzlibrary{patterns}

\usepackage{siunitx}
\usepackage{amsmath}
\usepackage{cases}
\usepackage{amsmath}
\usepackage{chemformula} 
\def\Ra{Ra}

\newcommand{\defn}{\stackrel{\text{def}}{=}}

\def\Racp{\mathcal{R}a_{cp}}
\def\Rar{\hat{R}a}

\newcommand*{\Eqdef}{\ensuremath{\mathrel{\overset{\mathrm{def}}{=}}}}

\shorttitle{Instability triggered by mixed convection}
\shortauthor{F.Rein, K.J.Burns, S.G.Llewellyn Smith, W.R.Young, B.Favier and M.Le Bars}

\title{Instability triggered by mixed convection in a thin fluid layer}

\author{
Florian Rein\aff{1,2}\corresp{\email{florian.rein@protonmail.com}}, 
Keaton J. Burns\aff{3,4},
Stefan G. Llewellyn Smith\aff{2,5}, 
William R. Young\aff{2}, 
Benjamin Favier\aff{1} and 
Michael Le Bars\aff{1}}

\affiliation{
\aff{1} Aix Marseille Universit\'e, CNRS, Centrale Med, IRPHE, Marseille, France 
\aff{2} Scripps Institution of Oceanography, University of California San Diego, La Jolla, CA 92093, USA
\aff{3} Department of Mathematics, Massachusetts Institute of Technology, Cambridge, MA 02139, USA
\aff{4} Center for Computational Astrophysics, Flatiron Institute, New York, NY 10010, USA
\aff{5} Department of Mechanical and Aerospace Engineering, Jacobs School of Engineering, University of
California San Diego, La Jolla, CA 92093, USA}

\begin{document}
\maketitle
\begin{abstract}
We investigate the convective stability of a thin, infinite fluid layer with a rectangular cross-section, subject to imposed heat fluxes at the top and bottom and fixed temperature along the vertical sides. 
The instability threshold depends on the Prandtl number as well as the normalized flux difference ($f$) and decreases with the aspect ratio ($\epsilon$), following a $\epsilon f^{-1}$ power law. 
Using 3D initial value and 2D eigenvalue calculations, we identify a dominant 3D mode characterized by two transverse standing waves attached to the domain edges.
We characterize the dominant mode's frequency and transverse wave number as functions of the Rayleigh number and aspect ratio. 
An analytical asymptotic solution for the base state in the bulk is obtained, valid over most of the domain and increasingly accurate for lower aspect ratios. 
A local stability analysis, based on the analytical base state, reveals oscillatory transverse instabilities consistent with the global instability characteristics.
The source term for this most unstable mode appears to be interactions between vertical shear and horizontal temperature gradients.
\end{abstract}


\section{Introduction}
\begin{figure}
   \centering
    \includegraphics[scale=0.25]{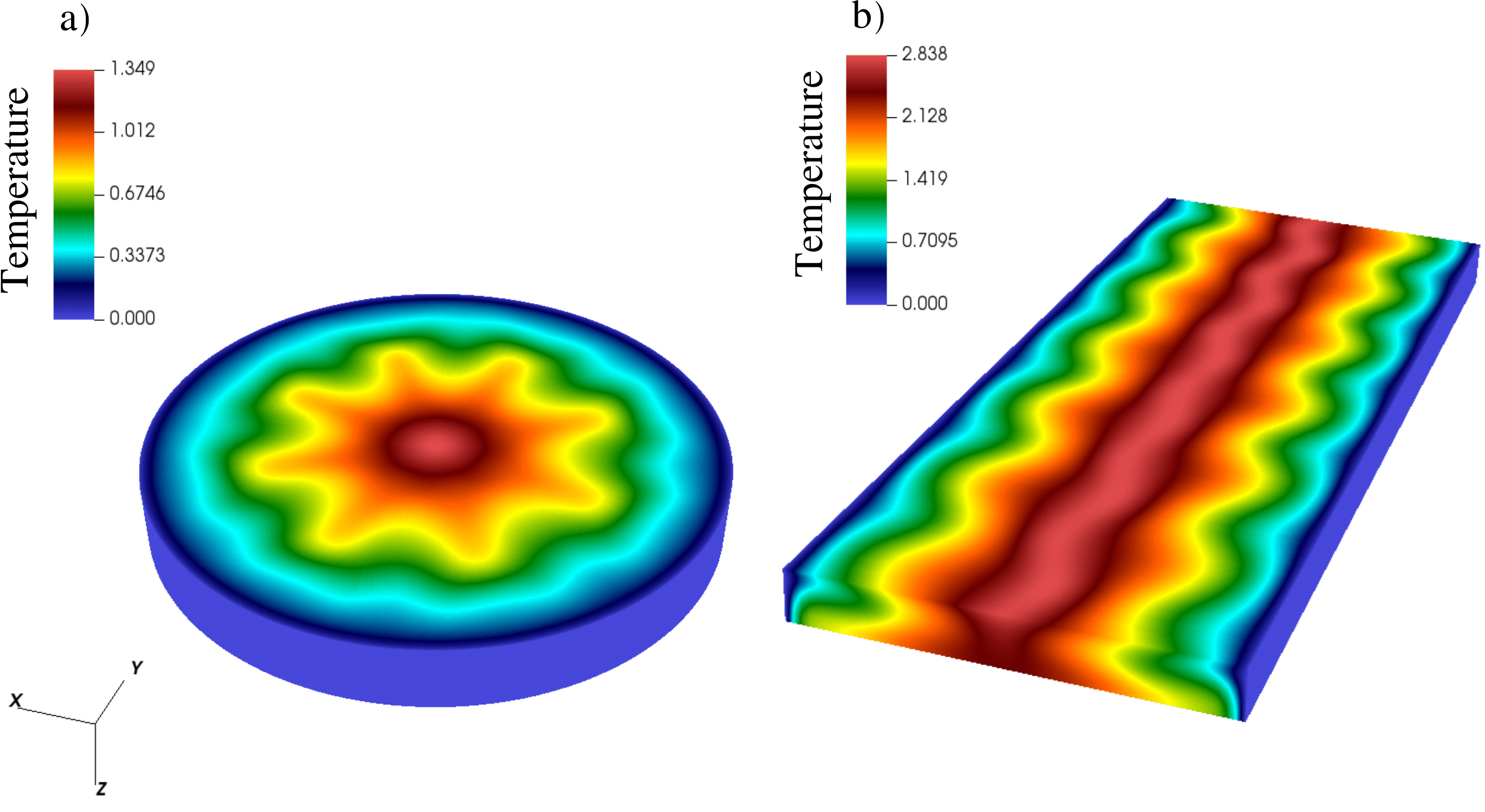}
   \caption{Snapshot of the bottom temperature field in a $(a)$ cylindrical with $\Ra=2\times10^4$ and $(b)$ Cartesian geometry with $\Ra=2\times10^3$. For both cases, the parameters are $\epsilon=1/4$, $Pr=0.1$ and $f=0.9$. To mimic the homogeneous azimuthal direction of the cylindrical case, periodic boundary conditions are applied in the transverse direction for the Cartesian case. Parameters and boundary conditions will be detailed in \autoref{sec:math}.}
   \label{fig1}
\end{figure}
The present work follows the numerical study of \cite{reinjfm} and the experimental study of \cite{reinmanip}, which was motivated by nuclear safety issues (see \cite{reinnureth}).
During a severe accident in a nuclear power plant, the radioactive fuel and metallic components of the reactor melt, forming a fluid known as corium. 
The corium moves from the reactor core to the lower plenum of the reactor vessel, where non-miscible oxidic and metallic phases separate. 
The oxide phase contains the majority of the radioactive elements and provides heat from below to the less dense and thinner liquid metal phase floating on its surface, which then focuses the heat towards the vessel wall. This phenomenon is referred to as the ``focusing effect'' in the nuclear safety literature. 
Understanding heat transfer through the top metal layer is essential for evaluating the risk of vessel failure and ensuring its integrity when ``in-vessel retention'' is employed as a severe accident management strategy \citep{THEOFANOUS,IVR}.

Our previous study of the turbulent metal layer dynamics \citep{reinjfm} used DNS of a thin cylindrical layer heated from below and cooled from above with imposed fluxes and cooled from the side with imposed temperature. It highlighted the presence of a three-dimensional (3D) drifting thermal structure composed of hot radial branches (see e.g.~\autoref{fig1}(a) which plays a crucial role in the spatiotemporal structure of heat transfer fluctuations. 
This drifting structure was further confirmed experimentally by \cite{reinmanip}.
The physical mechanisms behind the emergence of this pattern are so far unknown. 
It should be noted that this instability is not specific to the cylindrical geometry, but persists in Cartesian geometry, the focus of this paper.
\autoref{fig1} illutrates this point with two simulations performed using the same parameters but in cylindrical and Cartesian geometries.

In our system, a mixture of different types of convection can be expected.
Bottom heating and top cooling are reminiscent of Rayleigh-B\'enard configurations for which it is known that there exists a threshold below which diffusive processes dominate
over buoyancy forces, leading to a stable motionless base state \citep{chandrasekhar1981}.
The appearance and nonlinear dynamics of different patterns and structures above onset
have been extensively studied both theoretically \citep{FH_Busse_1978,Busse_Clever_1979, GOLUBITSKY1984} and experimentally \citep{Caldwell_1970,Krishnamurti1970OnTT,Bruyn1996}.
The case where the heat flux through the top is equal to the heat flux through the bottom was notably investigated by \cite{cp} and \cite{Shivashinsky}.
If the heat flux through the top is less than the heat flux through the bottom, then in steady state and in the absence of internal cooling, there must be an additional heat flux through the lateral boundaries (sidewalls).
This lateral cooling implies a horizontal temperature gradient which cannot be balanced by a hydrostatic pressure gradient. In this problem of natural convection the critical Rayleigh number is zero, i.e.~an arbitrarily small vertical flux inbalance induces convective motion.

Theoretical studies have examined the two-dimensional case in a closed box with insulated rigid walls at the top and bottom and differing imposed sidewall temperatures.
\cite{Cormack_Leal_Imberger_1974} considered small aspect ratio cases and showed, using matched asymptotic expansions, that the 2D flow consists of two distinct regimes: a parallel flow in the core region and a second non-parallel flow near the ends of the cavity.
A solution valid at all orders in the aspect ratio was found for the core region, while the first several terms of the appropriate asymptotic expansion were obtained for the end regions.
In the opposite asymptotic limit of a tall box, \cite{Gill_1966} found a 2D approximate solution for the velocity and temperature fields compatible with the vertically infinite solution of \cite{Batchelor}.
Low Prandtl regime dynamics has been investigated theoretically by \cite{HART1983}; a core solution, asymptotically valid at small aspect ratio was found, breaking down either as end effects extend into the center of the cavity or as a secondary shear flow instability develops in the core itself.
The stability of this flow was investigated by \cite{daniels1987} highlighting the role of the side boundary layer, as was further investigated by \cite{daniels_1993}.

The combination of these convection regimes referred  to as ``inclined'' or ``oblique'' temperature gradient was examined by \cite{Weber_1978}.
Both vertical (similar to Rayleigh-Bénard) and horizontal (similar to natural convection) temperature gradients are specified.
\cite{Weber_1978} considered the instability of the convective shear flow under the action of an inclined temperature gradient using a linear stability approach.
He found that varying the horizontal and vertical temperature gradients caused the flow to become unstable to different types of disturbances. 
These included longitudinal rolls aligned with the flow, transverse traveling rolls, and their oscillations.
\cite{ORTIZPEREZ2014,ORTIZPEREZ2015} studied the different instability modes that can appear, depending on the relative strength of the horizontal/vertical temperature gradients as well as the Prandtl number.
\cite{Patne_Oron_2022} and \cite{Dixit_Bukhari_Patne_2024} recently noted the destabilizing effect of horizontal temperature gradients leading to oblique oscillating modes for Prandtl number less than one.

All these results have been found for a fixed horizontal temperature gradient.
In our system, the horizontal temperature gradient is not prescribed but emerges dynamically from the mismatch between the heat fluxes at the top and bottom boundaries.
The stability of such a system and the physical mechanisms underlying the growth of instabilities are still unknown.
This paper presents a numerical and theoretical study of the instability onset in a thin, infinite fluid layer with a rectangular cross-section at low Prandtl number.
The objectives of this study are: (i) To find analytical solutions in the stable regime describing the temperature and velocity fields in the asymptotic limit of a small aspect ratio; 
(ii) To characterize the properties of the dominant mode at the onset of instability as a function of the parameters; and (iii) To identify the potential physical mechanisms related to the emergence of the thermal pattern observed in the turbulent regime by \cite{reinjfm,reinmanip}.

Section 2 outlines the governing equations and the numerical solvers employed. 
In section 3, we describe the base state by analyzing the flow structure and the different scalings involved in the low Rayleigh number regime, and propose an analytical asymptotic bulk solution.
In section 4, we characterize the first instability in the system by examining the structure of the dominant mode and how its properties change with the input parameters.
Additionally, we introduce a reduced model derived from the analytical asymptotic bulk solution, which successfully captures and reproduces the key characteristics of the instability.
Finally we discuss the implications of our findings to identify the potential physical mechanisms behind the instability. We conclude in Section 5.

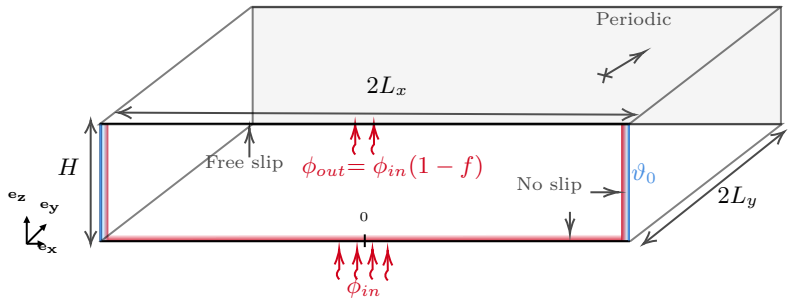
\begin{figure}
     \centering

  
\tikzset {_2wfvcobj7/.code = {\pgfsetadditionalshadetransform{ \pgftransformshift{\pgfpoint{0 bp } { -3.5 bp }  }  \pgftransformrotate{-269 }  \pgftransformscale{2 }  }}}
\pgfdeclarehorizontalshading{_1zjrrra6m}{150bp}{rgb(0bp)=(0.82,0.01,0.11);
rgb(37.5bp)=(0.82,0.01,0.11);
rgb(57.23214285714286bp)=(1,1,1);
rgb(100bp)=(1,1,1)}

  
\tikzset {_nlsuoo9ay/.code = {\pgfsetadditionalshadetransform{ \pgftransformshift{\pgfpoint{0 bp } { 0 bp }  }  \pgftransformrotate{0 }  \pgftransformscale{2.6 }  }}}
\pgfdeclarehorizontalshading{_jq4paqpxg}{150bp}{rgb(0bp)=(0.29,0.56,0.89);
rgb(37.5bp)=(0.29,0.56,0.89);
rgb(45.69940294538225bp)=(0.82,0.92,0.98);
rgb(62.5bp)=(0.82,0.01,0.11);
rgb(100bp)=(0.82,0.01,0.11)}

  
\tikzset {_fib9y84n0/.code = {\pgfsetadditionalshadetransform{ \pgftransformshift{\pgfpoint{0 bp } { 0 bp }  }  \pgftransformrotate{0 }  \pgftransformscale{2.6 }  }}}
\pgfdeclarehorizontalshading{_s46fs11us}{150bp}{rgb(0bp)=(0.82,0.01,0.11);
rgb(37.5bp)=(0.82,0.01,0.11);
rgb(53.39285714285714bp)=(0.82,0.92,0.98);
rgb(62.5bp)=(0.29,0.56,0.89);
rgb(100bp)=(0.29,0.56,0.89)}
\tikzset{every picture/.style={line width=0.75pt}} 

\begin{tikzpicture}[x=0.75pt,y=0.75pt,yscale=-1,xscale=1]

\draw  [draw opacity=0][shading=_1zjrrra6m,_2wfvcobj7][line width=0.75]  (297.28,14189.86) -- (559.46,14189.86) -- (559.46,14194.74) -- (297.28,14194.74) -- cycle ;
\draw  [draw opacity=0][shading=_jq4paqpxg,_nlsuoo9ay][line width=0.75]  (297.07,14194.75) -- (297.07,14136.05) -- (300.93,14136.04) -- (300.93,14194.73) -- cycle ;
\draw  [draw opacity=0][shading=_s46fs11us,_fib9y84n0][line width=0.75]  (558.17,14194.75) -- (558.17,14136.06) -- (562.03,14136.05) -- (562.03,14194.74) -- cycle ;
\draw [color={rgb, 255:red, 74; green, 74; blue, 74 }  ,draw opacity=1 ]   (308,14130.51) -- (563.12,14131.52) ;
\draw [shift={(565.12,14131.53)}, rotate = 180.23] [color={rgb, 255:red, 74; green, 74; blue, 74 }  ,draw opacity=1 ][line width=0.75]    (6.56,-2.94) .. controls (4.17,-1.38) and (1.99,-0.4) .. (0,0) .. controls (1.99,0.4) and (4.17,1.38) .. (6.56,2.94)   ;
\draw [shift={(306,14130.5)}, rotate = 0.23] [color={rgb, 255:red, 74; green, 74; blue, 74 }  ,draw opacity=1 ][line width=0.75]    (6.56,-2.94) .. controls (4.17,-1.38) and (1.99,-0.4) .. (0,0) .. controls (1.99,0.4) and (4.17,1.38) .. (6.56,2.94)   ;
\draw [color={rgb, 255:red, 74; green, 74; blue, 74 }  ,draw opacity=1 ]   (292.06,14194.28) -- (292.06,14136.32) ;
\draw [shift={(292.06,14134.32)}, rotate = 90] [color={rgb, 255:red, 74; green, 74; blue, 74 }  ,draw opacity=1 ][line width=0.75]    (6.56,-2.94) .. controls (4.17,-1.38) and (1.99,-0.4) .. (0,0) .. controls (1.99,0.4) and (4.17,1.38) .. (6.56,2.94)   ;
\draw [shift={(292.06,14196.28)}, rotate = 270] [color={rgb, 255:red, 74; green, 74; blue, 74 }  ,draw opacity=1 ][line width=0.75]    (6.56,-2.94) .. controls (4.17,-1.38) and (1.99,-0.4) .. (0,0) .. controls (1.99,0.4) and (4.17,1.38) .. (6.56,2.94)   ;
\draw    (266,14196) -- (260,14196) ;
\draw [shift={(269,14196)}, rotate = 180] [fill={rgb, 255:red, 0; green, 0; blue, 0 }  ][line width=0.08]  [draw opacity=0] (3.57,-1.72) -- (0,0) -- (3.57,1.72) -- cycle    ;
\draw    (260,14196.25) -- (260,14186.5) -- (260,14185) ;
\draw [shift={(260,14182)}, rotate = 90] [fill={rgb, 255:red, 0; green, 0; blue, 0 }  ][line width=0.08]  [draw opacity=0] (3.57,-1.72) -- (0,0) -- (3.57,1.72) -- cycle    ;
\draw    (267.88,14188.12) -- (260,14196) ;
\draw [shift={(270,14186)}, rotate = 135] [fill={rgb, 255:red, 0; green, 0; blue, 0 }  ][line width=0.08]  [draw opacity=0] (3.57,-1.72) -- (0,0) -- (3.57,1.72) -- cycle    ;

\draw [color={rgb, 255:red, 208; green, 2; blue, 27 }  ,draw opacity=1 ]   (424.83,14151.42) .. controls (423.15,14149.77) and (423.14,14148.1) .. (424.79,14146.42) -- (424.79,14146.29) -- (424.73,14138.29) ;
\draw [shift={(424.72,14136.29)}, rotate = 89.59] [color={rgb, 255:red, 208; green, 2; blue, 27 }  ,draw opacity=1 ][line width=0.75]    (7.65,-2.3) .. controls (4.86,-0.97) and (2.31,-0.21) .. (0,0) .. controls (2.31,0.21) and (4.86,0.98) .. (7.65,2.3)   ;
\draw [color={rgb, 255:red, 74; green, 74; blue, 74 }  ,draw opacity=1 ]   (371.68,14152.86) -- (371.68,14140.19) ;
\draw [shift={(371.68,14138.19)}, rotate = 90] [color={rgb, 255:red, 74; green, 74; blue, 74 }  ,draw opacity=1 ][line width=0.75]    (7.65,-2.3) .. controls (4.86,-0.97) and (2.31,-0.21) .. (0,0) .. controls (2.31,0.21) and (4.86,0.98) .. (7.65,2.3)   ;
\draw [color={rgb, 255:red, 74; green, 74; blue, 74 }  ,draw opacity=1 ]   (532.45,14180.07) -- (532.45,14189.14) ;
\draw [shift={(532.45,14191.14)}, rotate = 270] [color={rgb, 255:red, 74; green, 74; blue, 74 }  ,draw opacity=1 ][line width=0.75]    (7.65,-2.3) .. controls (4.86,-0.97) and (2.31,-0.21) .. (0,0) .. controls (2.31,0.21) and (4.86,0.98) .. (7.65,2.3)   ;
\draw    (429.55,14191.32) -- (429.55,14197.68) ;
\draw [color={rgb, 255:red, 208; green, 2; blue, 27 }  ,draw opacity=1 ]   (434.22,14151.42) .. controls (432.54,14149.77) and (432.53,14148.1) .. (434.18,14146.42) -- (434.18,14146.29) -- (434.13,14138.29) ;
\draw [shift={(434.11,14136.29)}, rotate = 89.59] [color={rgb, 255:red, 208; green, 2; blue, 27 }  ,draw opacity=1 ][line width=0.75]    (7.65,-2.3) .. controls (4.86,-0.97) and (2.31,-0.21) .. (0,0) .. controls (2.31,0.21) and (4.86,0.98) .. (7.65,2.3)   ;
\draw [color={rgb, 255:red, 208; green, 2; blue, 27 }  ,draw opacity=1 ]   (416.8,14213.79) .. controls (415.12,14212.14) and (415.11,14210.47) .. (416.76,14208.79) -- (416.76,14208.66) -- (416.7,14200.66) ;
\draw [shift={(416.69,14198.66)}, rotate = 89.59] [color={rgb, 255:red, 208; green, 2; blue, 27 }  ,draw opacity=1 ][line width=0.75]    (7.65,-2.3) .. controls (4.86,-0.97) and (2.31,-0.21) .. (0,0) .. controls (2.31,0.21) and (4.86,0.98) .. (7.65,2.3)   ;
\draw [color={rgb, 255:red, 208; green, 2; blue, 27 }  ,draw opacity=1 ]   (425.8,14213.79) .. controls (424.12,14212.14) and (424.11,14210.47) .. (425.76,14208.79) -- (425.76,14208.66) -- (425.71,14200.66) ;
\draw [shift={(425.69,14198.66)}, rotate = 89.59] [color={rgb, 255:red, 208; green, 2; blue, 27 }  ,draw opacity=1 ][line width=0.75]    (7.65,-2.3) .. controls (4.86,-0.97) and (2.31,-0.21) .. (0,0) .. controls (2.31,0.21) and (4.86,0.98) .. (7.65,2.3)   ;
\draw [color={rgb, 255:red, 208; green, 2; blue, 27 }  ,draw opacity=1 ]   (433.52,14213.79) .. controls (431.84,14212.14) and (431.83,14210.47) .. (433.48,14208.79) -- (433.48,14208.66) -- (433.42,14200.66) ;
\draw [shift={(433.41,14198.66)}, rotate = 89.59] [color={rgb, 255:red, 208; green, 2; blue, 27 }  ,draw opacity=1 ][line width=0.75]    (7.65,-2.3) .. controls (4.86,-0.97) and (2.31,-0.21) .. (0,0) .. controls (2.31,0.21) and (4.86,0.98) .. (7.65,2.3)   ;
\draw [color={rgb, 255:red, 208; green, 2; blue, 27 }  ,draw opacity=1 ]   (441.13,14214.32) .. controls (439.45,14212.67) and (439.44,14211) .. (441.09,14209.32) -- (441.09,14209.19) -- (441.03,14201.19) ;
\draw [shift={(441.02,14199.19)}, rotate = 89.59] [color={rgb, 255:red, 208; green, 2; blue, 27 }  ,draw opacity=1 ][line width=0.75]    (7.65,-2.3) .. controls (4.86,-0.97) and (2.31,-0.21) .. (0,0) .. controls (2.31,0.21) and (4.86,0.98) .. (7.65,2.3)   ;
\draw [color={rgb, 255:red, 74; green, 74; blue, 74 }  ,draw opacity=1 ]   (542.73,14170.29) -- (554.88,14170.29) ;
\draw [shift={(556.88,14170.29)}, rotate = 180] [color={rgb, 255:red, 74; green, 74; blue, 74 }  ,draw opacity=1 ][line width=0.75]    (7.65,-2.3) .. controls (4.86,-0.97) and (2.31,-0.21) .. (0,0) .. controls (2.31,0.21) and (4.86,0.98) .. (7.65,2.3)   ;
\draw [color={rgb, 255:red, 74; green, 144; blue, 226 }  ,draw opacity=1 ][line width=0.75]    (297.07,14136.05) -- (297.07,14194.74) ;
\draw  [line width=0.75]  (297.07,14136.05) -- (562.03,14136.05) -- (562.03,14194.75) -- (297.07,14194.75) -- cycle ;
\draw [color={rgb, 255:red, 74; green, 144; blue, 226 }  ,draw opacity=1 ][line width=0.75]    (562.03,14136.05) -- (562.03,14194.74) ;
\draw [color={rgb, 255:red, 74; green, 144; blue, 226 }  ,draw opacity=1 ][line width=0.75]    (297.07,14136.06) -- (297.07,14194.75) ;
\draw  [color={rgb, 255:red, 0; green, 0; blue, 0 }  ,draw opacity=0.51 ][fill={rgb, 255:red, 155; green, 155; blue, 155 }  ,fill opacity=0.1 ][line width=0.75]  (373.32,14077.67) -- (638.28,14077.67) -- (638.28,14136.37) -- (373.32,14136.37) -- cycle ;
\draw [color={rgb, 255:red, 74; green, 74; blue, 74 }  ,draw opacity=1 ]   (550,14112) -- (568.29,14101.03) ;
\draw [shift={(570,14100)}, rotate = 149.04] [color={rgb, 255:red, 74; green, 74; blue, 74 }  ,draw opacity=1 ][line width=0.75]    (7.65,-2.3) .. controls (4.86,-0.97) and (2.31,-0.21) .. (0,0) .. controls (2.31,0.21) and (4.86,0.98) .. (7.65,2.3)   ;
\draw [shift={(550,14112)}, rotate = 14.04] [color={rgb, 255:red, 74; green, 74; blue, 74 }  ,draw opacity=1 ][line width=0.75]    (-3.91,0) -- (3.91,0)(0,3.91) -- (0,-3.91)   ;
\draw  [color={rgb, 255:red, 0; green, 0; blue, 0 }  ,draw opacity=0.51 ][line width=0.75]  (372.48,14077.88) -- (638.03,14077.88) -- (562,14136.37) -- (296.46,14136.37) -- cycle ;
\draw  [color={rgb, 255:red, 0; green, 0; blue, 0 }  ,draw opacity=0.5 ] (373.33,14136.15) -- (638.87,14136.15) -- (562.85,14194.63) -- (297.3,14194.63) -- cycle ;
\draw [color={rgb, 255:red, 74; green, 74; blue, 74 }  ,draw opacity=1 ]   (570.69,14194.98) -- (638.43,14141.24) ;
\draw [shift={(640,14140)}, rotate = 141.58] [color={rgb, 255:red, 74; green, 74; blue, 74 }  ,draw opacity=1 ][line width=0.75]    (6.56,-2.94) .. controls (4.17,-1.38) and (1.99,-0.4) .. (0,0) .. controls (1.99,0.4) and (4.17,1.38) .. (6.56,2.94)   ;
\draw [shift={(569.13,14196.22)}, rotate = 321.58] [color={rgb, 255:red, 74; green, 74; blue, 74 }  ,draw opacity=1 ][line width=0.75]    (6.56,-2.94) .. controls (4.17,-1.38) and (1.99,-0.4) .. (0,0) .. controls (1.99,0.4) and (4.17,1.38) .. (6.56,2.94)   ;

\draw (544,14079) node [anchor=north west][inner sep=0.75pt]  [font=\scriptsize,color={rgb, 255:red, 74; green, 74; blue, 74 }  ,opacity=1 ] [align=left] {Periodic};
\draw (605,14166) node [anchor=north west][inner sep=0.75pt]  [font=\footnotesize]  {$2L_{y}$};
\draw (504.14,14161.42) node [anchor=north west][inner sep=0.75pt]   [align=left] {{\scriptsize \textcolor[rgb]{0.29,0.29,0.29}{No slip}}};
\draw (347.98,14150.45) node [anchor=north west][inner sep=0.75pt]   [align=left] {{\scriptsize \textcolor[rgb]{0.29,0.29,0.29}{Free slip}}};
\draw (418.84,14211.17) node [anchor=north west][inner sep=0.75pt]  [font=\footnotesize]  {$\textcolor[rgb]{0.82,0.01,0.11}{\phi }\textcolor[rgb]{0.82,0.01,0.11}{_{in}}$};
\draw (425.41,14178.97) node [anchor=north west][inner sep=0.75pt]  [font=\tiny]  {$0$};
\draw (396.41,14150.57) node [anchor=north west][inner sep=0.75pt]  [font=\footnotesize]  {$\textcolor[rgb]{0.82,0.01,0.11}{\phi }\textcolor[rgb]{0.82,0.01,0.11}{_{out}}\textcolor[rgb]{0.82,0.01,0.11}{=\phi }\textcolor[rgb]{0.82,0.01,0.11}{_{in}}\textcolor[rgb]{0.82,0.01,0.11}{{(1-f)}}$};
\draw (561.82,14154.53) node [anchor=north west][inner sep=0.75pt]  [font=\footnotesize,color={rgb, 255:red, 74; green, 144; blue, 226 }  ,opacity=1 ]  {$\vartheta _{0}$};
\draw (274,14152) node [anchor=north west][inner sep=0.75pt]  [font=\footnotesize]  {$H$};
\draw (429,14111) node [anchor=north west][inner sep=0.75pt]  [font=\footnotesize]  {$2L_{x}$};
\draw (265,14174) node [anchor=north west][inner sep=0.75pt]  [font=\tiny]  {$\mathbf{e_{y}}$};
\draw (249,14169) node [anchor=north west][inner sep=0.75pt]  [font=\tiny]  {$\mathbf{e_{z}}$};
\draw (264,14195) node [anchor=north west][inner sep=0.75pt]  [font=\tiny]  {$\mathbf{e_{x}}$};

\end{tikzpicture}

     \caption{Sketch of the modeled fluid metal layer inside the box. Differing top and bottom fluxes are imposed while the sidewall temperatures are fixed and uniform.}
    \label{intro}
\end{figure}

\section{Mathematical and numerical formulation}\label{sec:math}
\subsection{Governing equations}
We consider the flow of an incompressible fluid under the Boussinesq approximation.
The fluid is confined within a box of thickness $H$, length $2L_x$, periodic in the $y$-direction with period $2L_y$ (see Figure \ref{intro}) and with gravity pointing downward: $\bm{g}=-g\bm{e}_z$.
The fluid is heated from below with a uniform heat flux per unit area $\phi_{\textrm{in}}$ and cooled from above by a uniform outgoing flux $\phi_{\textrm{out}}$.
Note that we are interested in the cases where $\phi_{\textrm{in}}>\phi_{\textrm{out}}$ so that the residual heat flux necessarily escapes the domain through the sidewall boundaries at $x = \pm L_x$.
The dimensional temperature on the $x$-side wall boundaries is fixed at $\vartheta_0$.
We use no-slip boundary conditions at the bottom and lateral boundaries, and free-slip at the top boundary.
Using a no-slip upper boundary condition does not qualitatively change the results discussed below.
Lengths are rescaled using the height of the box $H$, while time is rescaled using the vertical diffusive timescale $H^2/\kappa$, with $\kappa$ the constant thermal diffusivity.
The dimensionless temperature $T$ is defined relative to the imposed side temperature and rescaled using the imposed bottom flux $\phi_{\textrm{in}}$, so that
\begin{equation}
\label{ndt}
    T=\frac{ k_{\text{th}}}{\phi_{\textrm{in}}H}\left(\vartheta-\vartheta_0\right) \ ,
\end{equation}
where $k_{\text{th}}$ is the thermal conductivity.
The dimensionless conservation equations for momentum, mass and energy are then
\begin{equation}
\begin{aligned}
Pr^{-1}\left(\bm{u}_t+ \bm{u\cdot\nabla}\bm{u}\right)=- \bm{\nabla}P +\Ra  T\bm{e}_z+ \bm{\nabla}^2\bm{u} \ ,
\end{aligned}
\label{qdm}
\end{equation}
\begin{equation}
\bm{\nabla\cdot u}=0 \ ,
\label{m}
\end{equation}
\begin{equation}
T_t+ \bm{u\cdot\nabla}T=\nabla^2 T \ .
\label{nrj}
\end{equation}
Here and from now on, $A_\alpha$ denote the partial derivative of $A$ with respect to the variable $\alpha$ and $\bm{u}$, $P$ and $T$ are the dimensionless velocity, pressure and temperature of the fluid.
The dimensionless boundary conditions can be written as

\begin{equation}
\begin{aligned}
\bm{u}=\bm{0} \quad \mbox{and}~~T_z=-1 \quad \mbox{at} \quad z=0 \ ,\\
u_{z}=v_{z}=w=0 \quad \mbox{and} \quad T_{z}=-1 + f \quad \mbox{at} \quad z=1\ ,\\
\bm{u}=\bm{0} \quad \mbox{and} \quad  T=0 \quad \mbox{at} \quad x=\epsilon^{-1} \ ,\\
\end{aligned}
\label{bcn}
\end{equation}
where $\bm{u}=\left(u,v,w\right)$ are the velocity components.
The problem is characterized by four main dimensionless parameters:
the aspect ratio $\epsilon$, the normalized flux difference $f$, the Rayleigh number $\Ra$ based on the heat flux imposed at the bottom $\phi_{\textrm{in}}$, and the Prandtl number $Pr$. They are defined by 
\begin{equation}
\epsilon = \frac{H}{L_x},\quad f=1-\frac{\phi_{\textrm{out}}}{\phi_{\textrm{in}}}, \quad \Ra=\frac{\beta g \phi_{\textrm{in}} H^4}{k \nu \kappa}, \quad Pr=\frac{\nu}{\kappa} \ ,
\end{equation}
where $\beta$ is the thermal expansion coefficient and $\nu$ is the kinematic viscosity, both assumed to be constant.
Added to this is the ratio $L_y/H$, fixed to 40 throughout the paper, which is large enough to ensure the natural development of the instability free from confinement effects along
the $y$-direction.
The present study explores the parameter space over a wide range, with $\epsilon$ varying from $1/8$ to $1/1024$, $\Ra$ from $3\times10^{-2}$ to $1.5\times10^4$, the normalized flux difference spanning $0.05 \leq f \leq 0.9$ and $Pr$ from 0.1 to 1.0. The lower bound $Pr=0.1$ is chosen to represent metallic mixtures of zirconium, uranium oxide, and steel relevant to nuclear applications, for which $Pr$ lies in the range 0.07–0.2 \citep{pr}, in line with previous studies.
A complete list of the simulations performed is provided in \autoref{A}.

The governing equations \eqref{qdm}--\eqref{nrj} with the boundary conditions \eqref{bcn} are solved using two primary methods. 
First, Direct Numerical Simulations (DNS) are employed to solve both 3D Initial Value Problems (IVPs), starting from given initial conditions and following their evolution in time, and 2D Boundary Value Problems (BVPs), which determine $y$-invariant steady-state solutions by solving \eqref{qdm}-\eqref{nrj} after removing temporal and $y$-derivatives.
Second, an Eigenvalue Problem (EVP) approach is used to analyze the stability of these stationary solutions. Details of all methods are provided in \autoref{appendixB}.

\subsection{Numerical approach}
Two different numerical codes were used in this study.
Dedalus is employed to determine the 2D base flows and temperature fields for each configuration via steady nonlinear boundary value problems (BVPs) and to investigate the stability of the system by solving global and local eigenvalue problems (EVPs). 
Nek5000 is used to perform 3D direct numerical simulations, solving initial value problems (IVPs) for given configurations.

\subsubsection{Nek5000}
The governing equations \eqref{qdm}--\eqref{nrj} with boundary conditions \eqref{bcn} and periodic boundary condition in the $y$ direction, are solved using the spectral element code \href{https://nek5000.mcs.anl.gov/}{Nek5000} \citep{Fischer1997,Deville2002} (IVP). 
The Cartesian geometry is discretized with up to $\mathcal{E}=4096$ hexahedral elements, refined near all boundaries to resolve viscous and thermal boundary layers accurately.
Numerical simulations are initiated with a quiescent fluid and a uniform temperature field $T=0$ throughout the domain. 
Small random temperature perturbations of amplitude $10^{-3}$ are introduced, leading to thermal convection growth during a transient phase lasting $\sim$ $10$ vertical diffusive times, with longer durations observed for lower aspect ratios ($\epsilon$). 
In order to test numerical convergence, we gradually increase the spectral polynomial order until convergence on the viscous dissipation is reached, as further described in \cite{reinjfm}.

\subsubsection{Dedalus}
Background and stability calculations are conducting using Dedalus \citep{dedalus}, a parallel pseudo-spectral solver.
Our model uses a high-order continuous spectral element method on a 2D domain $(x,z) \in [0, L_x] \times [0,1]$ to solve the $y$-invariant governing equations \eqref{qdm}--\eqref{nrj} with boundary conditions \eqref{bcn} and parity conditions imposed at $x=0$.
All fields are represented by bivariate Chebyshev expansions on a series of rectangular elements that expand dyadically away from the boundary as $[0, 0.5L_x], [0.5L_x, 0.75L_x], ... [L_x-4,L_x-2],[L_x-2,L_x-1],[L_x-1,L_x]$.
Each element employed up to $N_x \times N_z = 20\times 20$ spectral modes and $3/2$ dealiasing.
This discretization is used to solve nonlinear boundary value problems (BVP) to determine steady base solutions via the Newton-Kantorivich method.

Global 3D eigenvalue problems (Global-EVP) are solved with the same discretization using the Dedalus eigentools extension \citep{Oishi2021}. 
These problems use a fixed set of parameters: $\Ra$, $Pr$, $\epsilon$ and $f$ across a range of $k_y$ values from $10^{-2}$ to $5 \times 10^1$. 
The same approach is applied to 2D local eigenvalue problems (Local-EVP), using the base solution's vertical structure at a particular $x$ as a uniform background, with the same parameters, but now across a range of $k_y$ values from $0$ to $2$ and $k_x$ values from $-2$ to $2$.
To ensure convergence, we verify that the eigenmode and eigenvalue of the most unstable case remain consistent even when the spectral resolution is doubled.
For the sake of consistency, we compared the growth rates and mode structures from IVP simulations with Nek5000 against those obtained from the Global-EVP analysis using Dedalus.

A summary of all the numerical results involving each specific method, IVP, BVP, Global-EVP and Local-EVP, is given in \autoref{table:1} from \autoref{A}. 

\section{Base flow analysis}\label{sec:baseflow}
\subsection{Qualitative overview}
We begin with a qualitative description of the stationary two-dimensional base flow in the $(\bm{e_x},\bm{e_z})$ plane, obtained from IVP at steady state
and for a very low $Ra_\phi=10$, well below the instability threshold discussed later.
We have checked that exactly the same results are obtained using the BVP approach using Dedalus.
First, regardless of the control parameters, a motionless steady state does not exist in this system. 
Imposing different fluxes at the bottom and top boundaries generates a horizontal temperature gradient that cannot be balanced by a hydrostatic pressure gradient.
This results in natural convection.
Maintaining a constant temperature at the sides further reinforces this natural convection via a downward recirculation along the side boundary and a return flow toward the centre along the bottom boundary.

\begin{figure}
   \centering
    \includegraphics[scale=0.3]{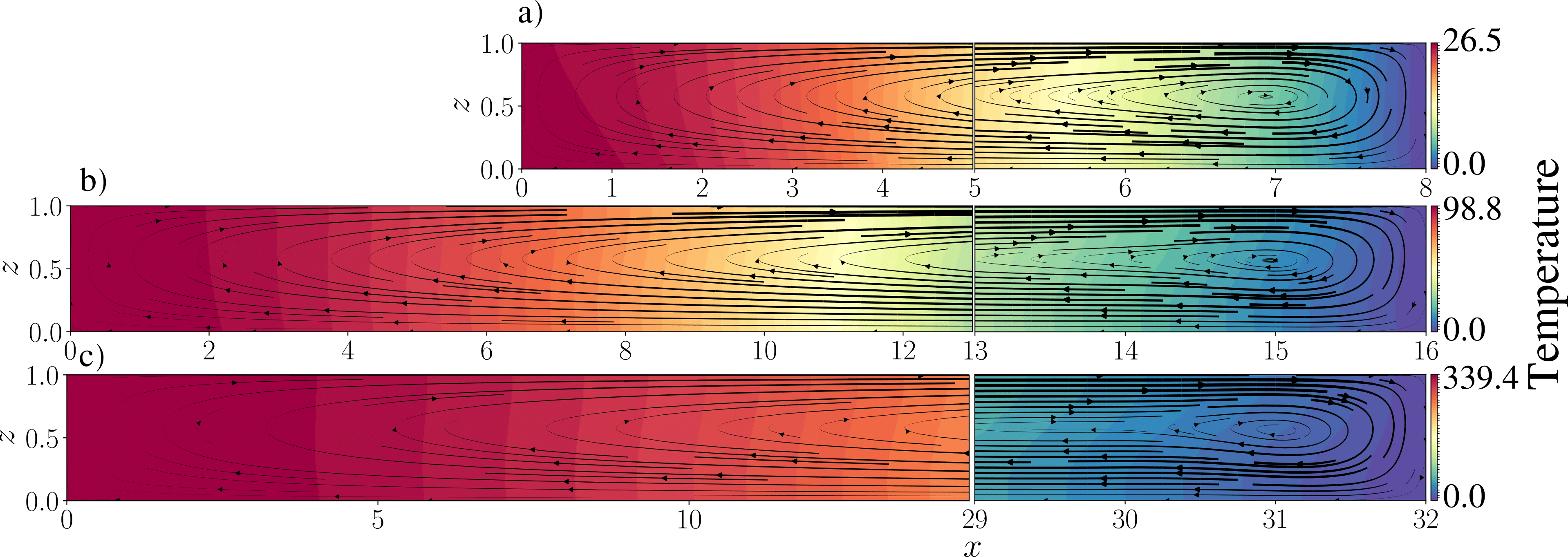}
   \caption{ 2D view in a plane ($x,z$) of the temperature
together with some velocity streamlines computed for aspect ratio ($a$) 1/8, ($b$) 1/16, ($c$) 1/32.
For each aspect ratio, a close-up view spanning three dimensionless units along the side wall is highlighted.
The IVP method is used to obtain the data, which corresponds to the simulations IVP1 to IVP3 listed in \autoref{table:1}. The parameters are $f=0.9$, $Pr=0.1$, $\Ra=10$.}
   \label{fig3}
\end{figure}

When temperatures are fixed on the top and bottom boundaries, the global recirculation covering the entire domain disappears, leaving only a vortex structure near the edge, as observed by \cite{GANZAROLLI}. 
This vortex structure, however, remains when heat fluxes are imposed at the top and bottom boundaries, extending over a penetration length similar to the layer thickness and independent of the aspect ratio, as shown in \autoref{fig3}.
At low $\Ra$ (below the instability threshold discussed below), the flow and the temperature field are 2D, symmetric with respect to $x=0$, and invariant along the $y$-direction (or axisymmetric in the cylindrical case).
As the aspect ratio decreases, the influence of the side walls becomes confined to a smaller portion of the domain localized at the edges.

\subsection{Scalings}
\begin{figure}
   \centering
    \includegraphics[scale=0.3]{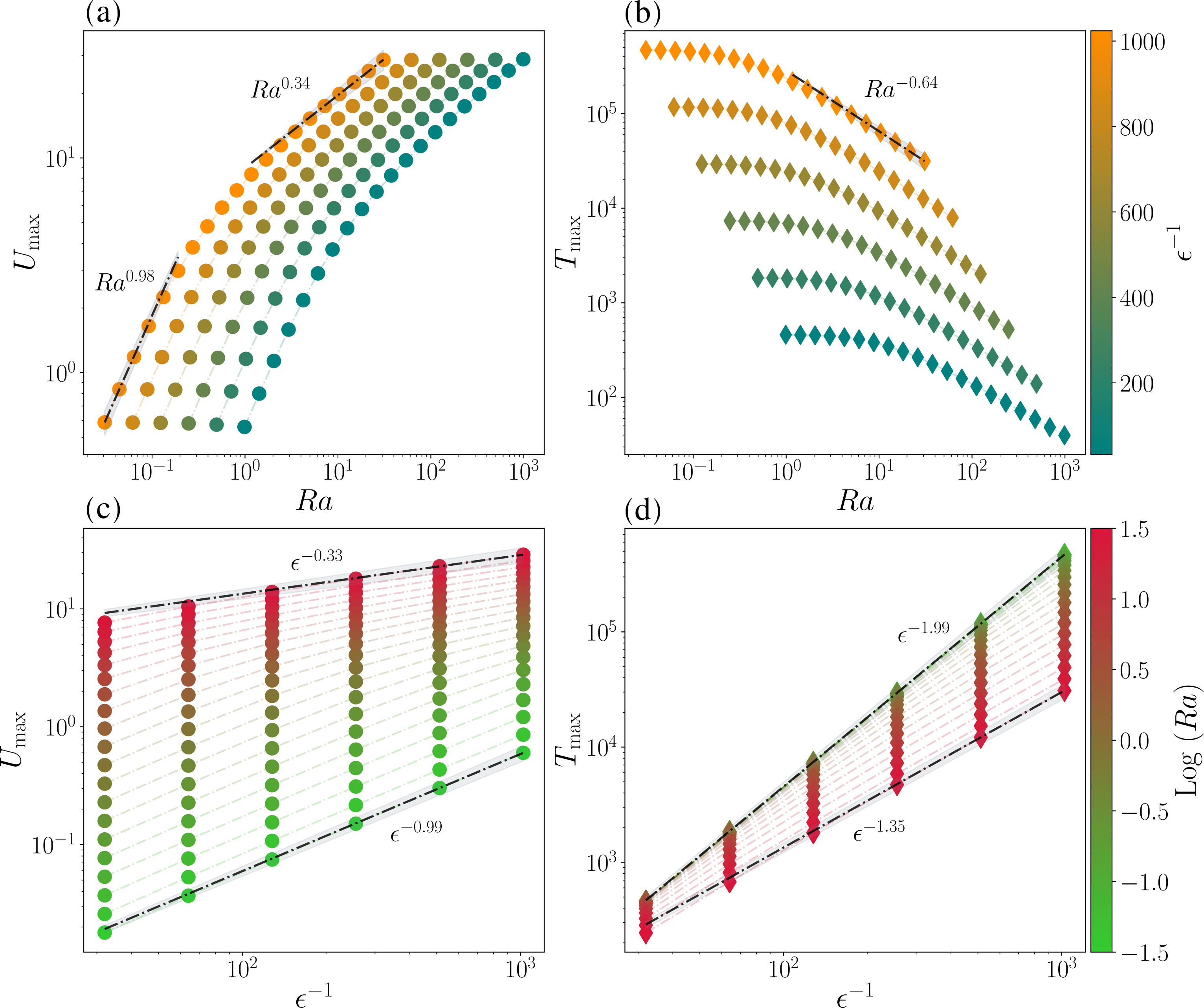}
   \caption{Peak values of $(a),(c)$ horizontal velocity, $(b),(d)$ temperature in the steady solution as a function of the Rayleigh number for aspect ratios from $1/1024$ to $1/32$ for $\Ra$ from $3\times 10^{-2}$ to $9.8\times10^2$ for $a)$, $b)$ and $3\times 10^{-2}$ to $3\times10^1$ for $c)$, $d)$. In each panel, the dash-dotted lines indicate the best-fit scaling law determined using a least squares method.
   The BVP method is used to obtain the data, corresponding to the simulations BVP1 to BVP6 listed in \autoref{table:1}.}
   \label{fig4}
\end{figure}

In this section we fix $Pr=0.1$ and $f=0.9$, and seek scaling laws for the maximal temperature and horizontal velocity of the 2D base solution denoted $T_{\text{max}}$ and $U_{\text{max}}$
respectively, systematically varying $\Ra$ and $\epsilon$.
For each simulation, the steady state is computed using a BVP approach with Dedalus.
The peak temperature is consistently located at the bottom centre, while the peak horizontal velocity is found at the top near the edge, just before the vortex recirculation area (see \autoref{fig3}).
The evolution of these peak values with $\Ra$ and $\epsilon$ is shown in \autoref{fig4}.

When the value of $\Ra$ increases at constant $\epsilon$, the peak horizontal velocity also increases, while the peak temperature decreases as expected (see \autoref{fig4}(a,b)). This occurs because a higher Rayleigh number enhances heat transfer efficiency within the system, causing the average temperature to approach the side temperature, which is zero in our dimensionless units \eqref{ndt}. 
Note, however, that the dimensional temperature increases with an increase in heat flux, $\Ra$.
We observe two main behaviors with $\Ra$: 
at relatively low $\Ra$, $T_{\text{max}}$ remains constant, while $U_{\text{max}}$ follows a power law behavior with an exponent unaffected by $\epsilon$.
Using a least squares method to estimate the power exponent, we find $U_{\text{max}}\sim Ra^{0.98\pm 0.01}$. 
To determine the mean exponent, we averaged the exponents for $\epsilon \in [1/32,1/1024]$ considering only data where $U_{\text{max}} \leq 2$, with variability quantified by the largest difference between these exponents.
At higher $\Ra$, both $U_{\text{max}}$ and $T_{\text{max}}$ follow a power law behavior in $\Ra$ with exponents unaffected by $\epsilon$.
Analyzing for each $\epsilon$ the 5 last $\Ra$ data points for $U_{\text{max}}$ and $T_{\text{max}}$, using a similar method, gives $U_{\text{max}}\sim \Ra^{0.34\pm 0.01}$ and $T_{\text{max}}\sim \Ra^{-0.64\pm 0.03}$. 

When $\epsilon$ decreases at constant $\Ra$, $U_{\text{max}}$ and $T_{\text{max}}$ increase (see \autoref{fig4}(c,d)).
Indeed, when $\epsilon$ decreases for a fixed $\Ra$, the ratio between the heating bottom surface and the cooling lateral surface also increases, leading to a larger global energy input into the system.  
We observe that a decrease in aspect ratio for a given $\Ra$ results in a greater horizontal temperature gradient (at least for non-turbulent flow).
Consequently, this enhanced gradient drives a more substantial horizontal flow, leading to a larger $U_{\text{max}}$.

We observe power-law behaviors for both $T_{\text{max}}$ and $U_{\text{max}}$ with exponents influenced by $\Ra$.
At lower $\Ra$, we estimate the power exponents using the least squares method and find $U_{\text{max}}\sim \epsilon^{-0.99\pm 0.01}$ and $T_{\text{max}}\sim \epsilon^{-1.99\pm 0.02}$.
To determine the mean exponent, we average the exponents, considering only data where $\Ra \leq 3 \times 10^{-1}$ and for $\epsilon \leq 1/128$, with variability quantified by the largest difference between these exponents.
At higher $\Ra$, analyzing data where $\Ra \geq 3\times 10^{-1}$ and for $\epsilon \leq 1/128$, using a similar methodology, we find  $U_{\text{max}}\sim \epsilon^{-0.33\pm 0.02}$ and $T_{\text{max}}\sim \epsilon^{-1.35\pm 0.02}$ .

\subsection{Theoretical analysis: Asymptotic bulk solution}\label{sec:anasol}
Motivated by observations from \autoref{fig3}, we are interested in cases where the aspect ratio $\epsilon$ is small, i.e.~a lubrication-type solution.
Our first goal is to compute the 2D steady-state solution for the bulk, stable for a sufficiently small Rayleigh number.
Let us detail the steady conservation equations that our analysis focuses on, which are also the equations solved by our BVP.
For a system invariant along the $y$-direction, the steady dimensionless conservation equations of momentum, mass and energy in the $x,z$ plane are
\begin{equation}
Pr^{-1}(U U_x + W U_{z}) =-P_x +  U_{xx} +  U_{zz},
\label{qdm2dU}
\end{equation}
\begin{equation}
    Pr^{-1}(U W_{x} + W W_{z}) =-P_{z} +\Ra T + W_{xx} + W_{zz},
    \label{qdm2dW}
\end{equation}
\begin{equation}
U_{x} + W_{z} =0 \ ,
\label{m2d}
\end{equation}
\begin{equation}
UT_{x} + WT_{z}=T_{xx} +  T_{zz}. \
\label{nrj2d}
\end{equation}
We define the vertical average by 
\begin{equation}
    \bar{A}(x) \Eqdef \int_0^1 \!\! A(x,z)~\mathrm{d}z.
\end{equation}
Integrating the continuity equation \eqref{m2d} over the depth of the fluid gives the exact result $\bar{U}_{x}=0$.
With impenetrable sidewalls, this implies $\bar{U}=0$.
We introduce the streamfunction $\psi$ such that  $(U, W) = (-\psi_{z},\psi_{x})$ and decompose the temperature field as 
\begin{equation}
T = T^b(z) + \tilde{T}(x,z), 
\end{equation}
with
\begin{equation}
T^b(z) = f(\tfrac{1}{2} z^2 - \tfrac{1}{6}) - z + \tfrac{1}{2}.
\label{boundaryTemp}
\end{equation}
The ``boundary temperature'' $T^b(z)$  is defined so that it satisfies the flux boundary conditions at $z=0$ and $1$ and has zero vertical average. 
Defining the rescaled horizontal coordinate $X=\epsilon x$, equations \eqref{qdm2dU}--\eqref{nrj2d} reduce to 
\begin{equation}
    Pr^{-1}\epsilon\mathcal{J}(\psi,\epsilon^2 \psi_{XX} + \psi_{zz}) = \epsilon\Ra ~\tilde{T}_X + \epsilon^4 \psi_{XXXX} + 2 \epsilon \psi_{zz XX} +  \psi_{zzzz},
    \label{vor1}
\end{equation}
\begin{equation}
    \epsilon\mathcal{J}(\psi,\tilde{T}) +\epsilon \psi_XT^b_z=\epsilon^2\tilde{T}_{XX} +\tilde{T}_{zz} + f,
    \label{nrj1}
\end{equation}
where  $\mathcal{J}$ is the Jacobian operator $\mathcal{J}(A,B) = A_{X}B_{z} - A_{z}B_{X}$.  The $y$-component of the vorticity is $ -\epsilon^2 \psi_{XX} - \psi_{zz}$.
\autoref{fig4} shows that in the diffusive regime ($\Ra \ll 1$)  the temperature scales as $T_{\mathrm{max}}\sim\epsilon^{-2}$. Thus we rescale the temperature by introducing rescaled temperature and Rayleigh numbers 
\begin{equation}\Theta = \epsilon^{2}\tilde{T}, \qquad \text{and} \qquad \Rar = \epsilon^{-1}\Ra.\label{scaling1}
\end{equation}
These scalings are appropriate in the lubrication limit $\epsilon \to 0$ and are further justified below.
With this notation equations \eqref{vor1}--\eqref{nrj1} are
\begin{align}
    Pr^{-1} \epsilon \mathcal{J}(\psi,\epsilon^2 \psi_{XX} + \psi_{zz}) &= \Rar~\Theta_X + \epsilon^4 \psi_{XXXX} + 2 \epsilon \psi_{XXzz} +  \psi_{zzzz},
    \label{asvor1}\\
    \epsilon \mathcal{J}(\psi,\Theta) +\epsilon^3\psi_XT^b_z &= \epsilon^2\Theta_{XX} + \Theta_{zz} + \epsilon^2f.
    \label{asnrj1}
\end{align}
Now integrate the energy equation \eqref{asnrj1} over depth.  The boundary temperature $T^b(z)$ is defined so that the vertical average of $\Theta_{zz}$ is zero, and thus one factor of $\epsilon$ can be canceled from this vertical average. Then an integration  in $X$ produces
\begin{equation}
\overline{\psi\Theta_z}+\epsilon^2 \overline{\psi T^b_z} + \epsilon \bar\Theta_{X} +\epsilon fX= 0.
    \label{HT}
\end{equation}
In equation \eqref{HT} the terms $\overline{\psi\Theta_z}$ and $\epsilon^2 \overline{\psi T^b_z}$ are the advective heat transport, while $\epsilon \overline{\Theta}_{X}$ is the horizontal diffusion of heat; the last term, $\epsilon fX$,  is the flux mismatch between the bottom and the top of the layer.

We seek a solution of \eqref{asvor1} and \eqref{asnrj1} for the streamfunction and the rescaled temperature expanded in powers of $\epsilon$ as
\begin{equation}
    \begin{aligned}
        \Theta(X,z)&= \Theta_0(X,z)+ \epsilon \Theta_1(X,z)  + \cdots,\\
        \psi(X,z) &= \psi_0(X,z) + \epsilon \psi_1(X,z) + \cdots~~.
    \end{aligned}
\end{equation}
Substituting these expansions into \eqref{asvor1} and \eqref{asnrj1} gives at the leading order
\begin{equation}
    {\Theta_0}_{zz} = 0,\qquad \qquad \text{and} \qquad \qquad
     {\psi_0}_{zzzz} = -\Rar{\Theta_0}_X.
     \label{0psi}
\end{equation}
The rescaled temperature has  vertical boundary conditions $\Theta_z(X,0) = \Theta_z(X,1)= 0$.  Thus  the solution of the temperature equation in \eqref{0psi} is $\Theta_0 = \bar\Theta_0(X)$ i.e.~the leading-order temperature is vertically uniform.
We can then solve the vorticity equation in  \eqref{0psi} to find 
\begin{equation}
    \psi_0 = \Rar \mathrm{P}(z)G(x), \qquad \qquad \text{where} \qquad \qquad G = \bar{\Theta}_{0X},
    \label{solpsi}
\end{equation}
and
\begin{equation}
    \mathrm{P}(z) = -\frac{z^4}{24} + \frac{5z^3}{48} - \frac{z^2}{16}.
    \label{P}
\end{equation}
$\mathrm{P}(z)$ is the polynomial equivalent to case c in \cite{cp}.
The temperature equation \eqref{asnrj1} at $O(\epsilon)$ is
\begin{equation}
    {\Theta_1}_{zz} = -{\psi_0}_z G,
    \label{1thetabalance}
\end{equation}
which  integrates in $z$ to 
\begin{equation}
    {\Theta_1}_{z} = -\Rar\mathrm{P}(z)G^2.
    \label{1theta}
\end{equation}
The construction above satisfies the boundary condition ${\Theta_1}_{z}=0$ at $z=0$ and $z=1$.

We do not need to compute explicitly $\psi_1$ to obtain $G$. Instead, consider the leading-order terms in the vertically averaged energy equation \eqref{HT} 
\begin{equation}
-\overline{{\psi_0} {\Theta_1}_z} +G\big(1- \epsilon \Rar \overline{\mathrm{P}T_z^b}\big) + fX = O(\epsilon).
    \label{1HT}
\end{equation}
Using equation \eqref{0psi} and \eqref{1theta} to evaluate the vertical averages in \eqref{1HT} in terms of the master variable $G(x)$ we find
\begin{equation}
    \alpha \Rar^2 G^3 + \left(1- \epsilon \Rar\frac{9 - 5 f}{2880}\right) G + fX = O(\epsilon),
    \label{cubic}
\end{equation}
where
\begin{equation}
    \alpha = \int_0^1 \mathrm{P}(z)^2~\mathrm{d}z = \frac{19}{1451520} \approx 1.3\times10^{-5}.
    \label{alpDef}
\end{equation}
The $G^3$-term  in \eqref{cubic} results from $\psi_0 \propto G$ and $\Theta_{1z} \propto G^2$ in \eqref{1HT}.

Below we proceed by discarding the $O(\epsilon)$ terms on the right of \eqref{cubic} and solving the resulting cubic equation for $G(X)$. We retain, however, the $O(\epsilon)$ term on the left  of \eqref{cubic}, i.e.~we retain the term involving $(9-5f)/2880$ in the coefficient of $G$. This heuristic step produces a small but palpable improvement in agreement between the predictions of \eqref{cubic} and the numerical results reported in \autoref{fig6} and \autoref{fig7}. 

Exploratory calculations indicate that the $O(\epsilon)$ term on the left of \eqref{cubic} has a different and simpler physical character than other $O(\epsilon)$ terms resulting from complicated expressions for $\psi_1$ and $\Theta_2$. In particular, let $\Racp(f)$ denote the critical Rayleigh number of the Rayleigh-Bénard instability with flux mismatch $f$;  \cite{cp} study the special case $f=0$. In  \autoref{Gcp} we generalize Chapman \& Proctor's linear stability calculation to non-zero $f$ and show that
\begin{equation}
\Racp(f) = \frac{2880}{9 - 5 f}.
\label{defRacp}
\end{equation}
$f=0$ recovers Chapman \& Proctor's critical Rayleigh number $320$. The $O(\epsilon)$ term on the left of \eqref{cubic} is therefore the ratio $\Ra/\Racp(f)$.

\subsection{Scaling}

The isolation of two-term dominant balances in the cubic equation \eqref{cubic}  helps us understand the scaling of velocity and temperature fields in low- and high-$\Rar$ asymptotic regimes.
With $\Rar \ll 1$ the cubic term $\alpha \Rar^2 G^3$ can be neglected,  leading to the balance $ G\approx -fX$ (for scaling arguments we also neglect the order $\epsilon$ term).
In this case the rescaled temperature profile that satisfies the side boundary condition ($\Theta_0=0$ at $X=1$)  is 
\begin{equation}
    {\bar\Theta_0}(X) \approx \frac{f}{2}\left(1-X^2\right).
\end{equation}
In this linear regime, diffusive processes dominate advection, and heat transport is mainly by diffusion. The  $\Rar \ll 1$ scaling for the temperature and horizontal velocity fields is
\begin{equation}
        \Theta \sim f,  \qquad \text{and} \qquad U \sim \Rar f.
        \label{lowRa}
\end{equation}
The rescaled temperature field is independent of $\Rar$ and $\epsilon$, while the horizontal velocity field is proportional to 
$\Rar$. These results match the scalings  in \autoref{fig4} where $T \sim \epsilon^{-2}\Theta \sim \epsilon^{-2}f $ and $U\sim \Ra \epsilon^{-1}f$ (see \eqref{scaling1}). 

Now consider the asymptotic regime $\alpha\Rar \gg 1$.
Then the $G^3$ term in \eqref{cubic} term dominates the middle term $G$, leading to the balance $  G\approx - (fX/\alpha \Rar^2)^{1/3}$.
In this case  the rescaled temperature profile is 
\begin{equation}
    {\bar\Theta_0}(X) \approx \frac{3}{4} (f/\alpha \Rar^2)^{1/3}\left(1-X^{4/3}\right).
\end{equation}
In the $\alpha\Rar \gg 1$ regime, the scalings are
\begin{equation}
        \Theta \sim \Rar^{-2/3} f^{1/3}, \qquad \text{and} \qquad
        U \sim \Rar^{1/3} f^{1/3}.
        \label{hiRa}
\end{equation}
These results further support the scalings shown in \autoref{fig4} as $T\sim \Ra^{-2/3}\epsilon^{-4/3}$ and $U\sim \Ra^{1/3}\epsilon^{-1/3}$.

At intermediate values of $\Rar$ we are in the transition regime between \eqref{lowRa} and \eqref{hiRa} i.e.~the three terms in the cubic \eqref{cubic} have the same order of magnitude
\begin{equation}
    \alpha \Rar^2 G^3 \sim  G \sim f.
\end{equation}
This three term balance implies that $\Rar$ is approximately equal to the ``regime transition Rayleigh number''
\begin{equation}
    \mathcal{R}\defn\frac{1}{\sqrt{\alpha}f}.
    \label{Ra_transition}
\end{equation}
\begin{figure}
   \centering
    \includegraphics[scale=0.3]{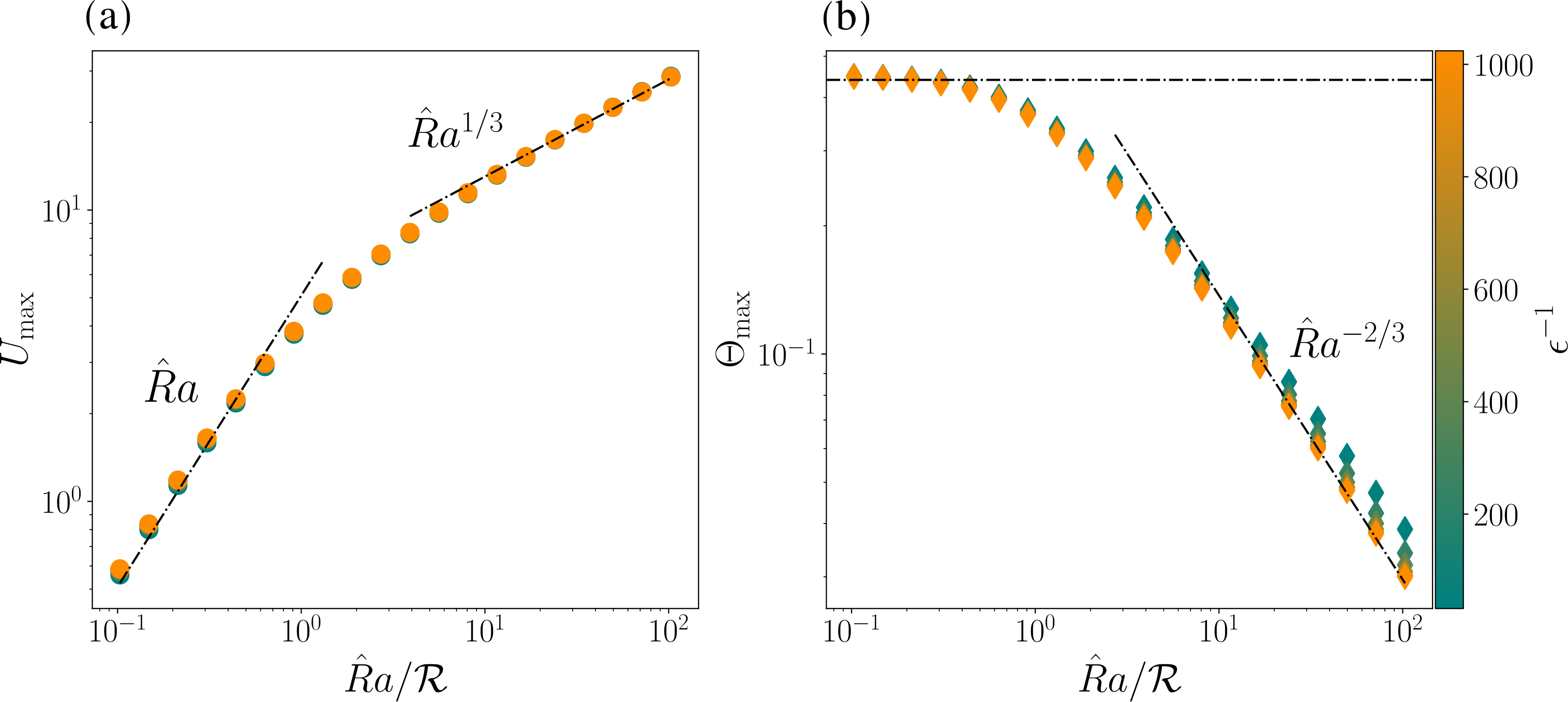}
   \caption{Peak values of $(a)$ horizontal velocity, $(b)$ rescaled maximal temperature $\Theta_{\mathrm{max}} = \epsilon^{-2} T_{\mathrm{max}}$ in the steady solution as a function of the rescaled Rayleigh number ($\Rar$) normalized by $\mathcal{R}$ from \eqref{Ra_transition} for several aspect ratios going from $1/1024$ to $1/32$. The BVP method is used to obtain the data, which corresponds to the simulations  BVP1 to BVP6 listed in \autoref{table:1}. The parameters are $Pr=0.1$ and $f=0.9$.}
   \label{fig5}
\end{figure}
Because $\alpha^{-1/2} \approx 276$,  $\mathcal{R}$ is significantly larger than $f^{-1}$. 
$\mathcal{R}$ corresponds to the Rayleigh number at which the advective transport of heat flux can no longer be neglected compared to diffusion. With $\Rar \sim \mathcal{R}$  we expect to see a change in scaling behavior, which transitions from the low Rayleigh regime to the high Rayleigh regime.
In \autoref{fig5}, we plot the data for the rescaled peak temperature and the horizontal peak velocity against $\Rar/\mathcal{R}$. 
At fixed $f$ the renormalized Rayleigh number $\Rar/\mathcal{R}$ is equivalent to dividing $\Ra$ by $\epsilon$. 
We see in \autoref{fig5} that with  this normalization, the data sets for different aspect ratios collapse onto the same curve for both horizontal velocity (\autoref{fig5}(a) and rescaled temperature data (\autoref{fig5}(b). 
\autoref{fig5} highlights the excellent agreement between the theoretically predicted scalings in the low- and high-Rayleigh asymptotic regimes and the numerical data.
The threshold at which the transition between these regimes occurs is consistent with the predictions from the scaling analysis: there is  a break in slope when $\Rar/\mathcal{R} \approx 1$.
\begin{figure}
   \centering
    \includegraphics[scale=0.3]{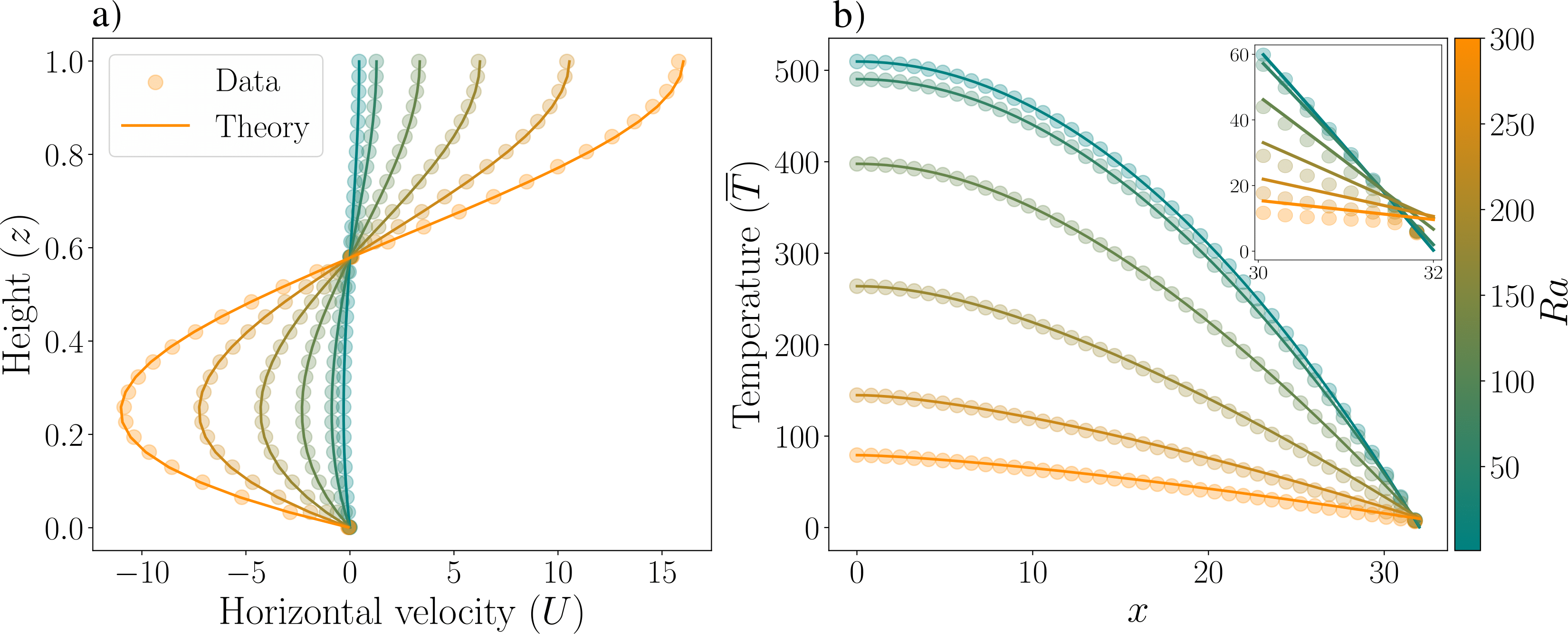}
   \caption{$(a)$ Vertical profile of the horizontal velocity at $X=2/3$, $(b)$ Horizontal temperature profile ($\overline{T}(x)$) for $\Ra$ from $1$ to $300$.
    The inset shows a zoom of $(b)$ in the $x \in [30;32]$. Circles ($\circ$) represent numerical data from IVP method, listed as IVP4 to IVP9 in \autoref{table:1}. Continuous lines are solutions of the cubic equation \eqref{cubic}. The parameters are $\epsilon=1/32$, $f=0.9$ and $Pr=0.1$.}
   \label{fig6}
\end{figure}

\begin{figure}
   \centering
    \includegraphics[scale=0.3]{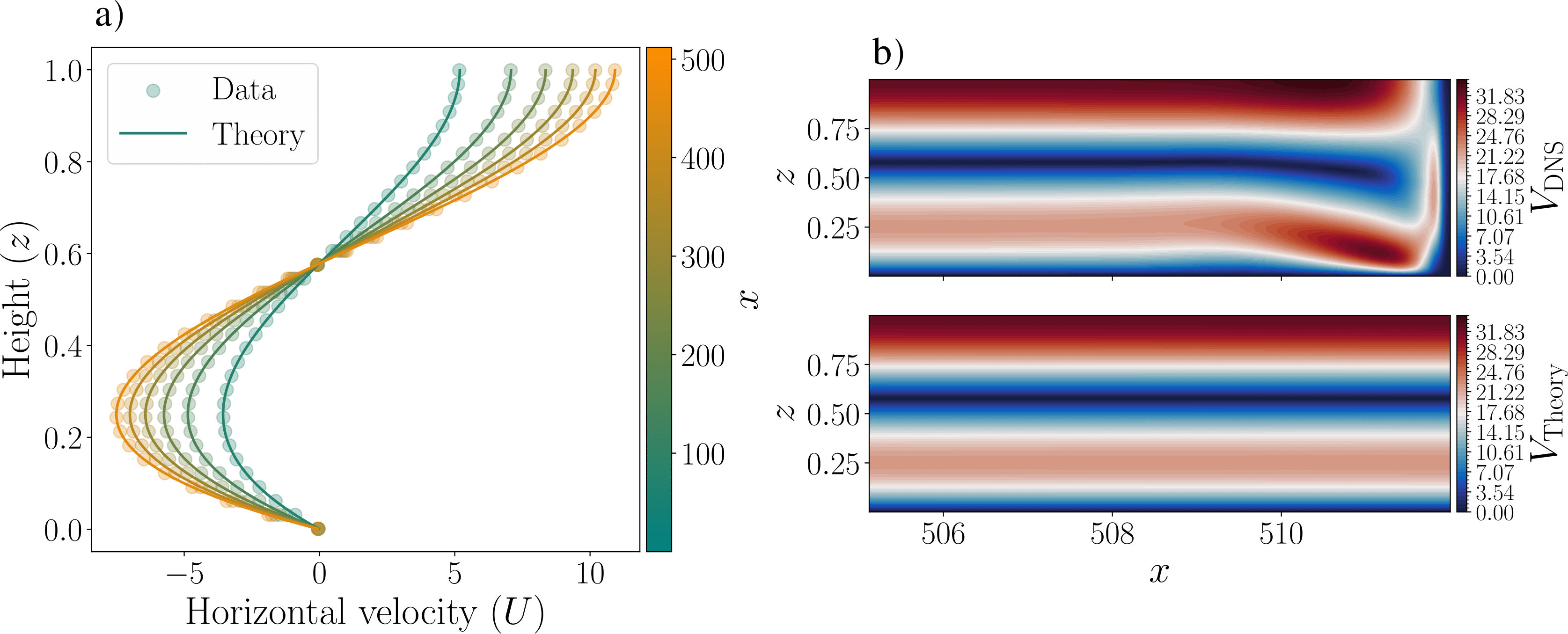}
   \caption{ $(a)$ Vertical profile of the base flow horizontal velocity ($U$) at several values of $x$ from $0$ to $510$.
    Circles ($\circ$) represent numerical data from IVP method denoted IVP10 in \autoref{table:1}, continuous lines are solutions from equation \eqref{solpsi}.
   $(b)$ Velocity magnitude in the $(x,z)$ plane: (top) numerical data from IVP10 ($V_{\mathrm{DNS}}$) and (bottom) analytical solution from equation \eqref{solpsi} ($V_{\mathrm{Theory}}$).}
   \label{fig7}
\end{figure}

\autoref{fig6} shows further excellent agreement between asymptotic solutions and numerical solutions. 
\autoref{fig6}a displays the vertical profile of horizontal velocity measured at $X = 2/3$, while \autoref{fig6}b shows the horizontal temperature profile for various values of $\Ra$.
The agreement between numerical and analytical solutions is robust across the entire $\Ra$ range. 
Discrepancies near the domain edges, corresponding to  recirculation zones, highlight the limitations in the theoretical solution.
These edge effects and the boundary conditions $T(X=\pm1,z)=0$ are not accounted for, becoming more pronounced as $\Ra$ increases.
The recirculation area, localized within a characteristic size comparable to the domain height of order 1 in dimensionless units (see e.g.~\autoref{fig3}), contributes to these discrepancies.
In \autoref{fig7}, the horizontal velocity profile for $\epsilon = 1/512$, $\Ra = 100$, $Pr = 0.1$, and $f = 1$ shows excellent agreement with the bulk solution up to $x = 510$ (\autoref{fig7}$a$).
\autoref{fig7}$b$ shows the velocity magnitude field in the $(x, z)$ plane compared to the numerical data from IVP method and the analytical solution from \eqref{solpsi}.
Discrepancies only emerge in the last two length units of the domain, coinciding with the recirculation zone.

In conclusion, the proposed analytical asymptotic solution exhibits very good agreement over a wide range of $\Ra$ values, covering a significant portion of the domain, particularly for small $\epsilon$. 
In the following section, we investigate the instabilities present within the system.

\section{Stability analysis}\label{stabanalysis}
In this section, we highlight the onset of the instability observed in \cite{reinjfm,reinmanip} and perform both IVP simulations and Global-EVP analysis to characterize it.
To understand the physical mechanisms behind this instability, we propose a reduced 1D model based on the analytical solution for the base state from the previous section. 
We perform a Local-EVP analysis on this reduced 1D model and characterize its stability.
We then analyse the similarities between the instabilities in the reduced 1D model and the global instability, allowing us to draw conclusions on the physical mechanism responsible for the latter.

\subsection{Phenomenology of the global instability}
Using IVP simulations from Nek5000 and Global-EVP from Dedalus, we characterize the first instability identified in the system.
We focus on identifying thresholds for different aspect ratio, normalized flux difference and Prandtl number, and evaluate the variation of the transverse wave number as well as angular frequency as a function of the input parameters.
\subsubsection{DNS analysis}
\begin{figure}
   \centering
    \includegraphics[scale=0.3]{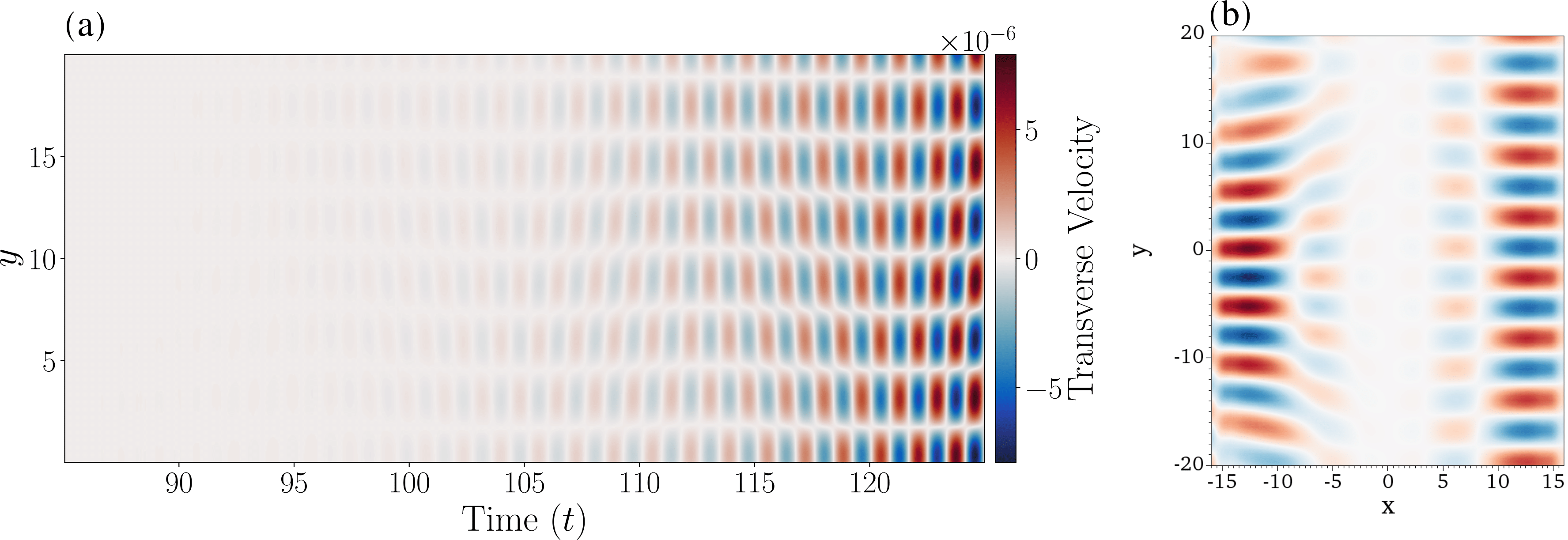}
   \caption{$(a)$ Hovm\"oller(space-time) diagram of the $z$-averaged transverse velocity component (along the periodic $y$-direction) measured at $X=3/4$. $(b)$ 2D snapshot in the $(x,y)$-plan of the transverse velocity at $t = 120$. The IVP-DNS method is used to obtain the data and corresponds to the IVP11 simulation listed in \autoref{table:1}. The parameters are $\epsilon=1/16$,$L_y=40$, $Pr=0.1$, $f=0.9$ and $\Ra=170$.}
   \label{fig8}
\end{figure}

We start the analysis by showing the transient dynamics of this system at a low value of $Ra_\phi=170$ (well below the typical values considered in \cite{reinjfm,reinmanip}) 
In \autoref{fig8}, we present a Hovm\"oller diagram showing the space-time evolution of the $z$-averaged transverse velocity ($v$), measured at $X=3/4$. 
\autoref{fig8} reveals the emergence of a three-dimensional instability near the sidewalls. 
For the given parameters, this instability occurs at a $\Ra$ value well below the Chapman \& Proctor purely thermal instability threshold of 640 (for $f=0.9$, see \autoref{Gcp}). 
The insets show that the instability is characterized by a transverse $y$-wavelength approximately seven times the layer height, a vertical wavelength equal to the layer height, and a horizontal structure extending six times the layer height in the $x$-direction, from the sides.
This horizontal structure decays as it moves toward the centre.
Two standing waves, positioned near each boundary in the $x$-direction and oscillating in antiphase, are observed.
Notice that an oscillating pattern is characteristic of convective instability at low Prandtl number \citep{Busse_1972,clever_1990}.

\subsubsection{Global EVP analysis}
We now use a global eigenvalue analysis to characterize the properties of the global instability observed at low $\Ra$ and low $Pr$.
\begin{figure}
   \centering
    \includegraphics[scale=0.35]{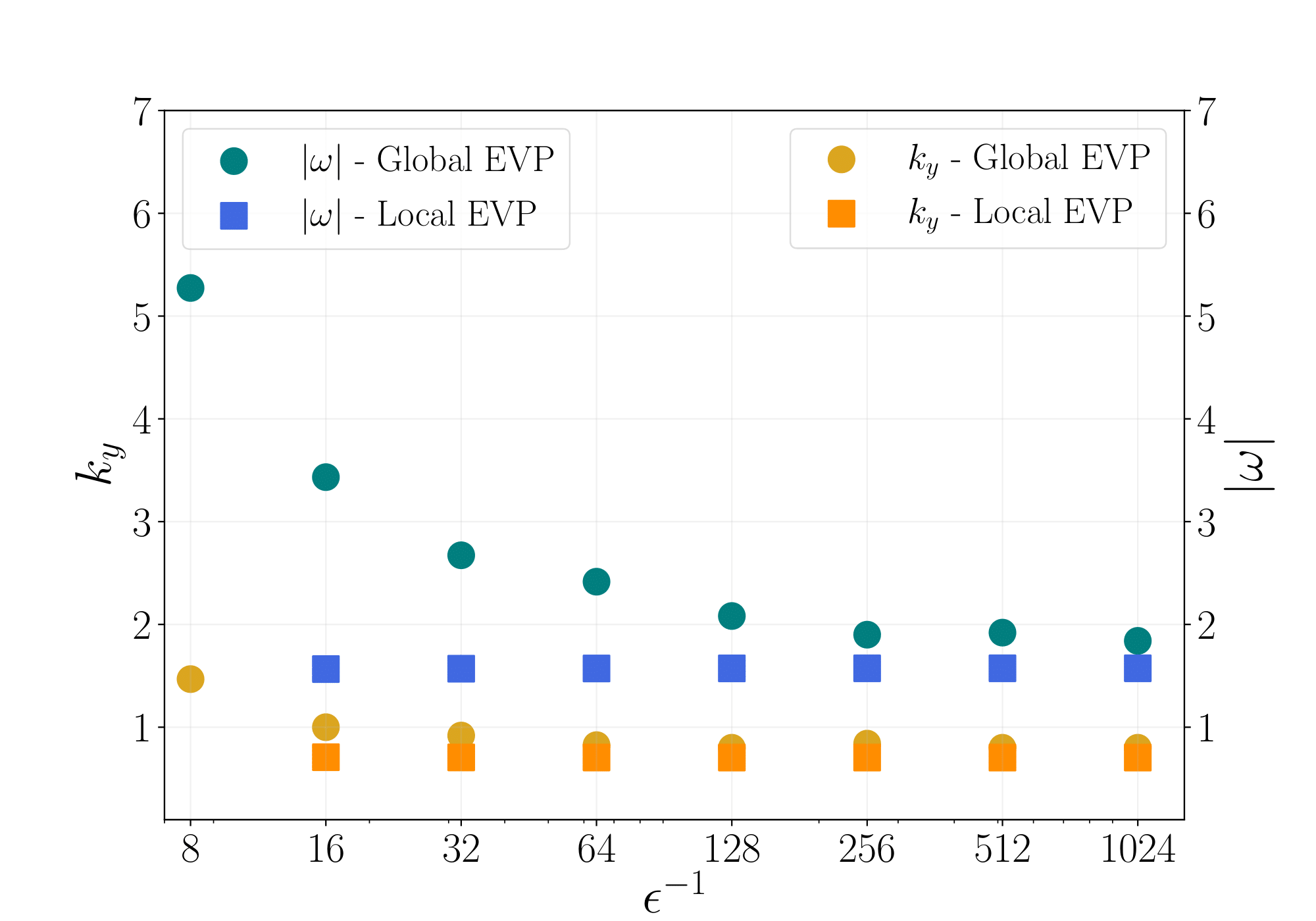}
   \caption{Angular frequency $|\omega|$ and transverse wavenumber $k_y$ as functions of $\epsilon$ for both Global and Local EVP analysis.
   For each method $\Ra$ is set to the threshold value identified for each aspect ratio. The Global-EVP method is used to obtain the data corresponding to simulations GEVP1 to GEVP5 and GEVP9 and the Local-EVP method is used to obtain the data corresponding to simulations LEVP1 to LEVP9 as listed in \autoref{table:1}. For both, $f=0.9$ and $Pr=0.1$. }
   \label{fig9}
\end{figure}

\autoref{fig9} shows the variation with $\epsilon$ of $|\omega|$ and $k_y$  of the first unstable mode at the critical value of $\Ra$.
The angular frequency $\omega$ is defined as the imaginary part of the complex growth rate $\sigma$.
Each plot focuses on the mode with the highest growth rate, measuring its transverse wavenumber and angular frequency.
In \autoref{fig9}, $\epsilon$ changes while $\Ra$ is fixed at the critical value for each aspect ratio.
At the onset of instability, as the aspect ratio decreases, the values of $k_y$ and $\omega$ tend to constant values, becoming independent of the aspect ratio.
This behaviour shows that the underlying physical mechanism driving the instability is  local.
\begin{figure}
   \centering
    \includegraphics[scale=0.32]{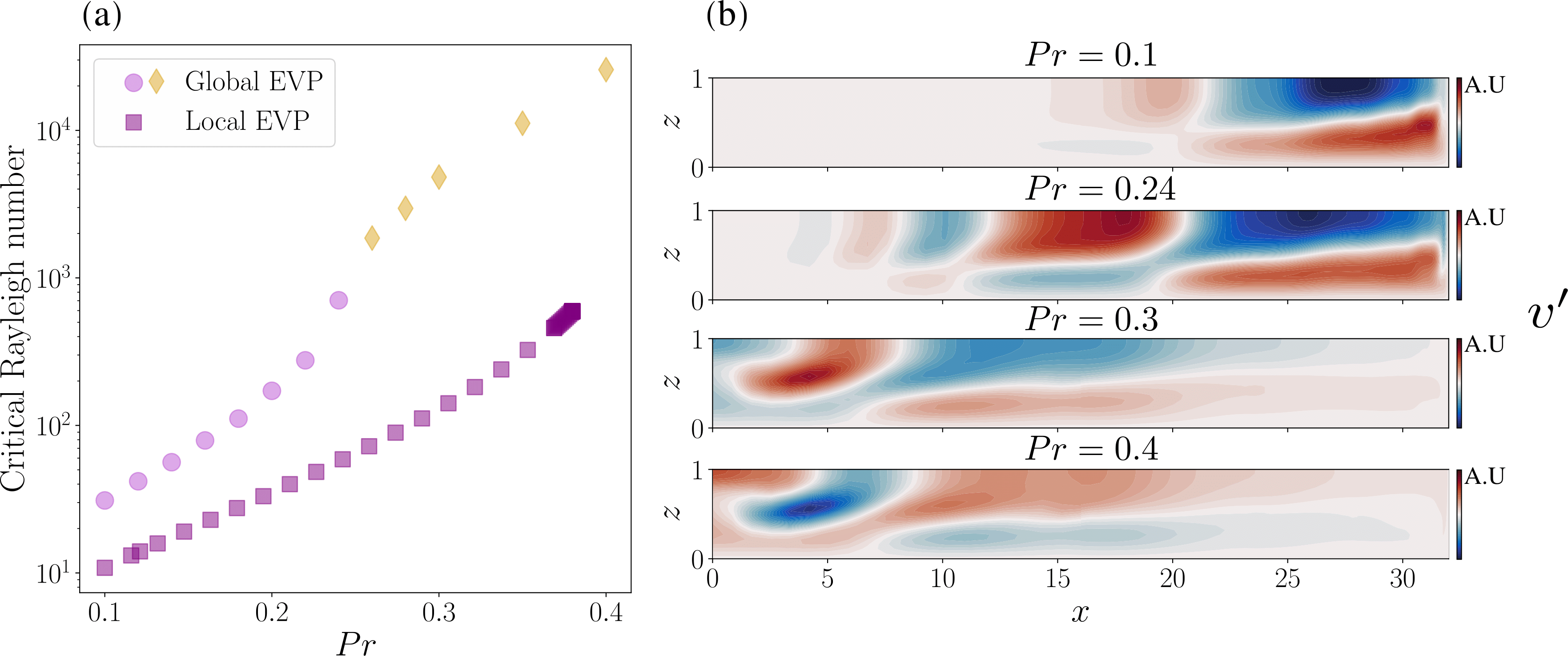}
   \caption{$(a)$ Critical Rayleigh number ($Ra_c$) as a function of the Prandtl number using both Global and Local EVP analysis. For all data, the critical wavenumber $k_y=0.92$ ($\circ$ and $\square$) except when $Pr \geq 0.26$ where $k_y=0.58$ ($\Diamond$). $(b)$ Perturbed transverse velocity for various Prandtl numbers from 0.1 at the top to 0.4 at the bottom at $\Ra=Ra_c$. The colorbars are centered around zero, with white indicating zero values. Positive values are shown in red and negative values in blue. The Global-EVP method is used to obtain the data corresponding to simulations GEVP10 to GEVP16, as listed in \autoref{table:1}.}
   \label{fig10}
\end{figure}

Next, we analyse the effect of the Prandtl number on the system stability. 
To do so, we consider the case $\epsilon=1/32$ and $f=0.9$.
For each Prandtl number from 0.1 to 0.4, we determine the critical Rayleigh number.
\autoref{fig10}(a) shows the critical Rayleigh number as a function of $Pr$.
We observe that increasing the Prandtl number has a stabilizing effect consistent with the qualitative observations in \autoref{fig18} (see Appendix \ref{StabCP}), where the system remains stable for $Pr = 1$ even at Rayleigh numbers exceeding the Chapman \& Proctor instability threshold.
The larger the Prandtl number the larger the threshold.
A slight change in behaviour can be noted when $Pr\simeq0.23$.
In \autoref{fig10}(b) we plot the eigenmode structure in a 2D $(x,z)$ plane of the perturbation of the transverse velocity for several Prandtl values at $\Ra=Ra_c$.
We observe an expanding structure of the mode in the $x$-direction when $Pr$ increases, until it covers the entire domain for $Pr=0.3$.
Close to the side, the structure of the mode changes when $Pr\approx 0.24$, corresponding to the slight change in behaviour of $Ra_c$ observed in \autoref{fig10}(b). 
The same expanding behaviour happens for all the other perturbation variables.
For $Pr < 0.24$, we find the same wave number $k_y=0.92$ associated with the most unstable mode.
However, when $Pr>0.24$ a different modal structure, mainly located at the center of the domain, is observed.
The critical wavenumber is found to be $k_y=0.58$.
The primary global instability mode, identified at low Prandtl number,  disappears between $0.2<Pr<0.3$, leading to another instability with much higher threshold.
 
\begin{figure}
   \centering
    \includegraphics[scale=0.35]{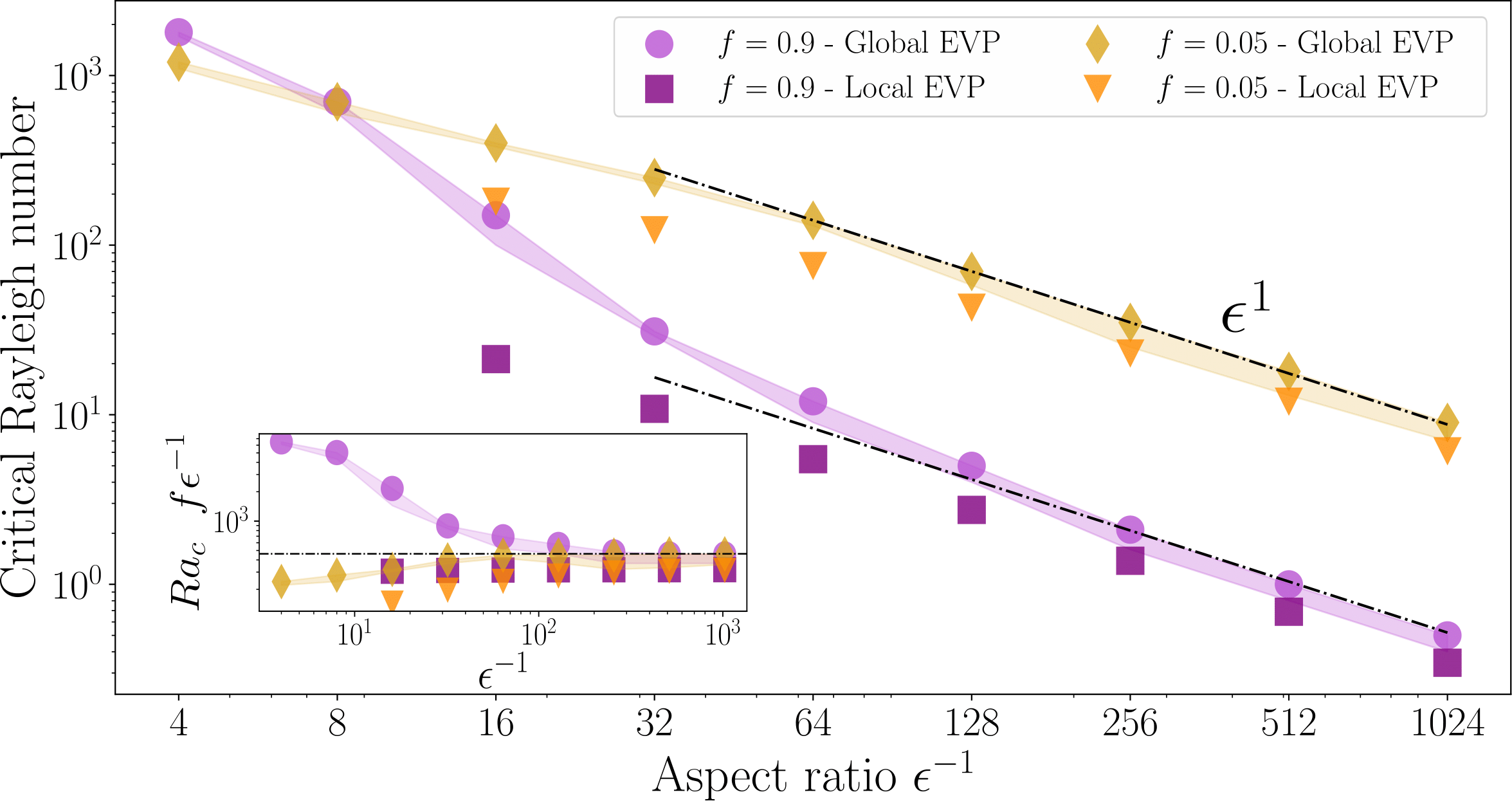}
   \caption{Critical Rayleigh number $Ra_c$ as a function of aspect ratio for $f=0.05$ and $f=0.9$ for both Global and Local EVP analysis. The dash-dotted lines indicate the best fit following a $\epsilon^{1}$ law.
   The inset shows the compensated critical Rayleigh number. The horizontal dash-dotted line is set at 462. The Global-EVP method is used to obtain the data corresponding to simulations GEVP1 to GEVP8, as listed in \autoref{table:1}. Data for $\epsilon=1/4$ are obtained using IVP-DNS simulation denoted IVP12. For all, $Pr=0.1$.}
   \label{fig11}
\end{figure}

To determine how the critical Rayleigh number ($Ra_c$) depends on the parameters, we conducted a stability analysis using the Global-EVP method.
Beginning with a starting $\Ra$ value, we varied $k_y$ from $10^{-2}$ to $5\times10^1$.
Employing a bisection method on $\Ra$, we  look for the first  $\Ra$ for all $k_y$ having a zero real part of the growth rate.
This method enabled us to approach the critical $\Ra$ as a function of $\epsilon$ and $Pr$. 
\autoref{fig11} shows that for $f=0.05$ and $f =0.9$, as $\epsilon$ decreases, the instability threshold decreases. 
Configurations with larger normalized flux differences are more unstable compared to those with lower normalized flux differences.
For both normalized flux difference regimes, a power law behaviour is noted, with a $+1$ exponent for the aspect ratio dependency.
The asymptotic behaviour is obtained at lower aspect ratio when $f$ is low.
For $f=0.05$ the asymptotic behaviour starts at $\epsilon\simeq1/64$, while we need to reach $\epsilon \simeq1/256$ for $f=0.9$.
In the inset, we plot the critical Rayleigh number compensated by $f\epsilon^{-1}$.
All the data collapse onto the same straight line indicating a universal law related to the global instability.
In the next section, we focus on identifying the physical mechanism underlying the global instability.

\subsection{Local analysis}
\label{localanalysis}
We have identified the characteristics of the instability as being three-dimensional, oscillatory, and emerging near the edge of the domain for low enough Prandtl number. 
To capture the physical mechanism responsible for this phenomenon, we study a simplified version of the problem.
\subsubsection{Reduced 1D model: dominant terms of the base state}
\begin{figure}
   \centering
    \includegraphics[scale=0.4]{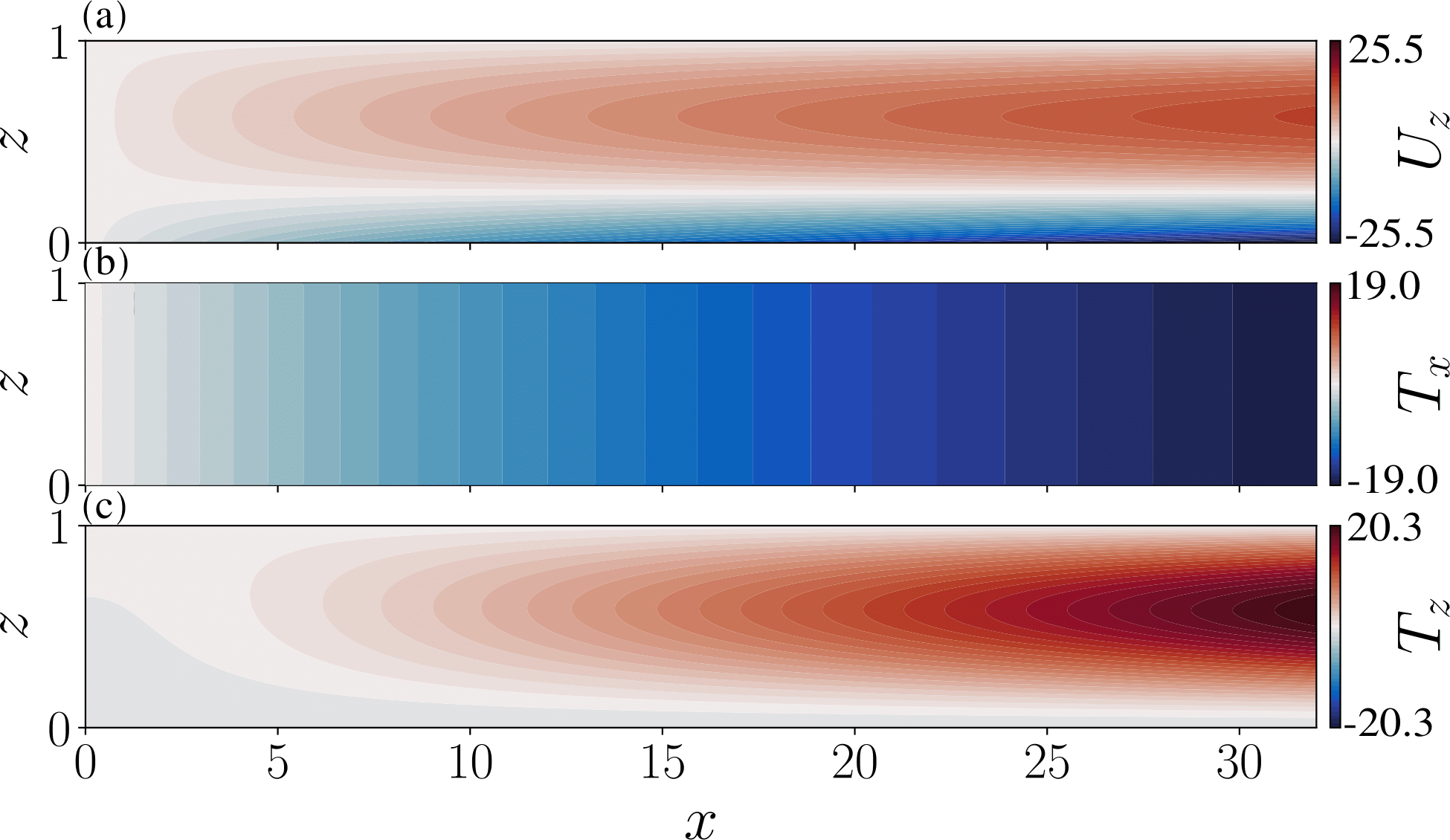}
   \caption{Simplified base state from the analytical solution \eqref{solpsi}, \eqref{1theta}, \eqref{cubic}. $(a)$ Vertical shear ($U_z$), $(b)$ horizontal and $(c)$ vertical  temperature gradient. The parameters are $Pr=0.1$, $f=0.9$, $\Ra=10.84$ and $\epsilon^{-1}=32$. }
   \label{fig12}
\end{figure}

To build a relevant reduced model, we aim to study the stability of the flow in the bulk of the system, without considering the impact of the sidewall.
To describe it, we use the steady-state analytical asymptotic solution of the base state detailed in \autoref{sec:anasol}.
We want to simplify the flow description by keeping only the dominant contributions.
We compute the horizontal temperature gradient $T_X = \epsilon^{-2} {\Theta_0}_X = \epsilon^{-2} G(x)$ using the leading-order rescaled temperature solution, solving the cubic equation \eqref{cubic}.
Velocities are computed from the streamfunction relation \eqref{solpsi} at the leading order and the vertical temperature gradient is computed using the $O(\epsilon)$ solution of the rescaled temperature thought the relationship
\begin{equation}
    T_z = T_z^b(z) + \epsilon^{-1}{\Theta_1}_z = fz - 1 - \epsilon^{-1} \Rar\mathrm{P}(z)G^2(x).
    \label{tzglobal}
\end{equation}
The mass conservation equation \eqref{m2d} implies that the vertical velocity is smaller by a factor of $\epsilon$ than the horizontal velocity.
Therefore, only the horizontal velocity is considered in describing the base flow.
We finally obtain a simplified base state description characterized by $\bm{U}=(U(z),0,0)$ with a  vertical shear $U_z$ as well as horizontal $T_x$ and vertical $T_z$ temperature gradients (see \autoref{fig12}).
As suggested by the location of the unstable mode in \autoref{fig8}, we perform a local analysis of the base state at $x = \epsilon^{-1}$.
This position coincides with the location where both velocity and thermal gradients are the strongest.

\subsubsection{Local EVP analysis}\label{LocalEVP}
We now perform a numerical eigenvalue analysis of the reduced 1D model, using the Local-EVP method described in \S \ref{EVPloc} solving the equations \eqref{ucart}--\eqref{Mcart} with the boundary conditions \eqref{bcnpertu}.
Note that contrary to the global EVP approach used above, which was homogeneous in $y$ only, the local EVP is homogeneous
in both $x$ and $y$-directions.

\begin{figure}
   \centering
    \includegraphics[scale=0.3]{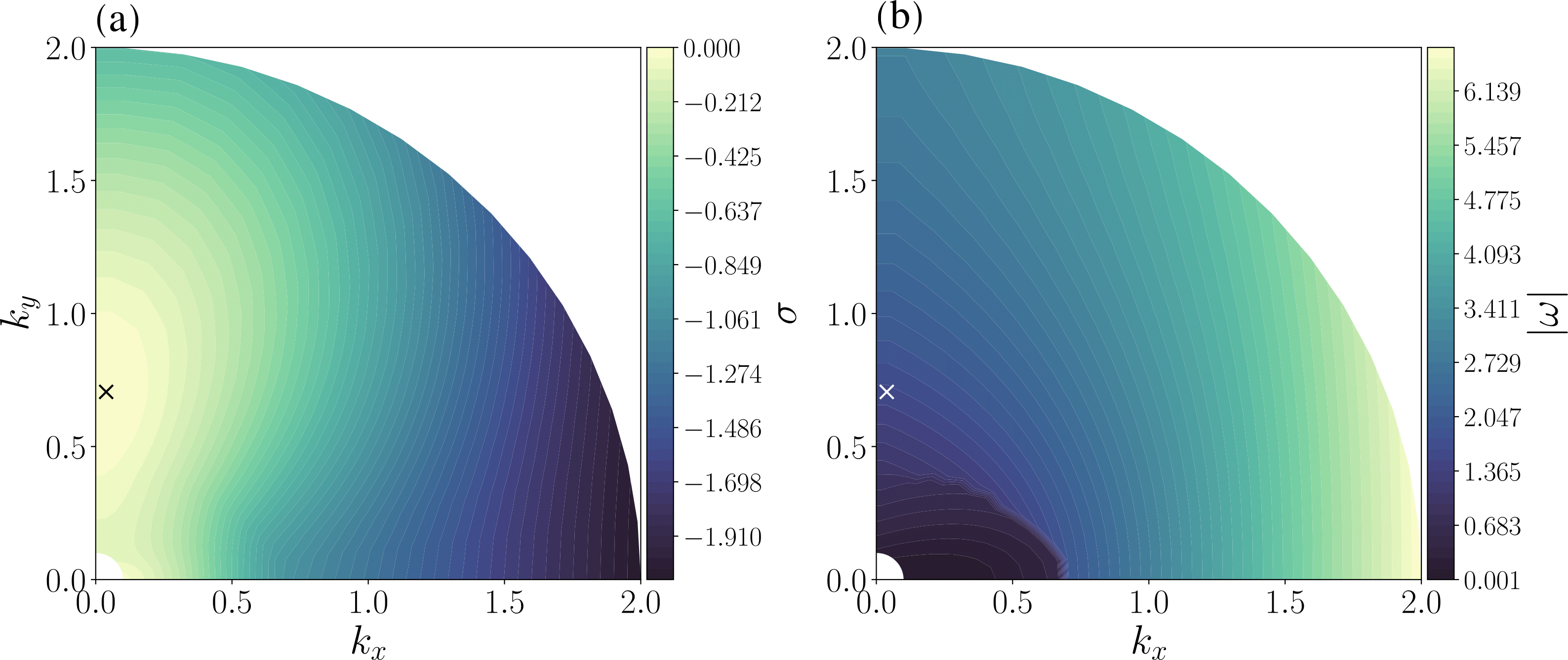}
   \caption{$(a)$ Real part of the growth rate and $(b)$ absolute value of the angular frequency in a $(k_x$, $k_y)$ map. The cross indicates the location of the maximal growth rate.}
   \label{fig13}
\end{figure}

The Local-EVP simulation listed in LEVP7 in \autoref{table:1} is performed for the case $x=1/128$, $Pr=0.1$, $f=0.9$ at just above the onset ($\sigma \gtrsim 0$) for $\Ra=2.75$.
\autoref{fig13} displays the growth rate and angular frequency in $(k_x,k_y)$ space for these parameters.
We observe a positive growth rate with the most unstable mode exhibiting transverse oscillations characterized by $k_y=7.0\times10^{-1}$, $k_x=1.93\times10^{-2}$, and $\omega=1.57$.
For $\epsilon=1/128$, the global instability is characterized by $Ra_c=5.0$ with $k_y=0.85$, $\omega=2.08$ for the most unstable mode.
These values are similar in magnitude to the results obtained from the Local-EVP method.
Despite the difference between the local ($Ra_c=2.75$) and global ($Ra_c=5.0$) thresholds, both remain significantly lower than the Chapman–Proctor threshold ($\Racp =640$).
The local analysis shows an oscillating 3D instability for the most unstable mode, but it is associated with $k_x\simeq 0$, revealing a limitation of our local approach.
In \autoref{fig9}, we present the values of $|\omega|$ and $k_y$ associated with the most unstable mode predicted by the local model, plotted as functions of $\epsilon$.
Across all aspect ratios, the local analysis consistently yields  $|\omega| = 1.57 $ and $k_y = 0.70 $.
As the aspect ratio decreases, the global analysis results for $|\omega|$ and $k_y$ converge toward these similar constant values obtained from the local model. 
The lower the aspect ratio, the better the agreement between the global and local analysis.

\begin{figure}
   \centering
    \includegraphics[scale=0.25]{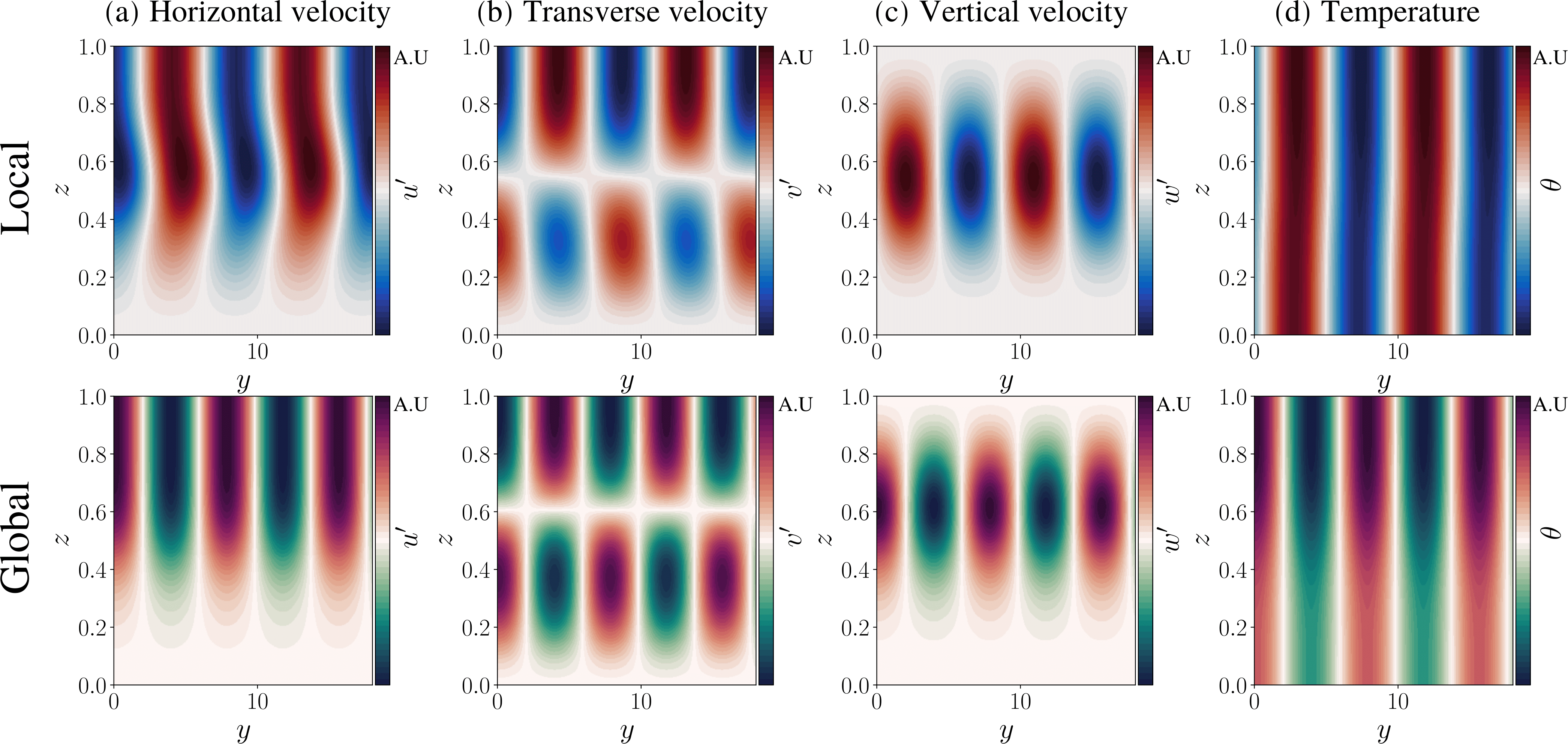}
   \caption{Eigenmode maps in a $(y,z)$ plane of the $(a)$ horizontal velocity, $(b)$ transverse velocity, $(c)$ vertical velocity and $(d)$ temperature. 
    The top row shows eigenmodes from local EVP computations, while the bottom row displays global EVP computations. The global eigenmodes are for $\Ra=5.0$, $\epsilon=1/128$ at $x=126$ with $k_y=0.85$ and 
    the local eigenmodes are for $\Ra=2.75$, $\epsilon=1/128$ at $x=128$ with $k_y=0.70$ and $k_x=1.93\times10^{-2}$. The colorbars are centered around zero, with white indicating zero values. Positive values are shown in red and negative values in blue
    Data correspond respectively to LEVP7 and GEVP5 simulations listed in \autoref{table:1}.} 
   \label{fig14}
\end{figure}

We now compare the vertical eigenmode structure for the most unstable mode from the Local-EVP analysis and from the Global-EVP analysis for the case $\epsilon=1/128$, $Pr=0.1$, $f=0.9$ and 
$\Ra=Ra_c$ ($Ra_c=2.75$ and 5.0 for the local and global analysis, respectively).
\autoref{fig14} illustrates the perturbation velocity and temperature fields maps in $(y,z)$ plane for the most unstable mode, taken at $x=126$ and 128 for the global and local analysis, respectively. 
We choose to look at the global mode structure at $x=126$ to avoid the influence of the recirculation flow near the edge.
There is a strong similarity in the vertical mode structure between the local and global approaches for all perturbation fields.

With this local approach, we investigate the Prandtl number dependence by solving Local-EVP at $x=32$, $f=0.9$, seeking the critical Rayleigh number for each Prandtl number value going from 0.1 to 0.38.
In \autoref{fig10}(a), we notice the same stabilizing effect for larger $Pr$ as that identified in the global analysis:
the larger $Pr$, the larger the threshold.
It should be noted that the most unstable mode reaches an asymptotic value $Pr\simeq 0.38$ and then disappears (i.e.~a mode with a completely different structure, present but subcritical before, now becomes the most unstable one). 
This behaviour is consistent with the changing mode structure identified for $Pr\approx0.24$ from the global analysis.
This transition in mode structure also explains why, at larger Prandtl numbers (e.g. $Pr = 0.3$), the critical Rayleigh number from the global EVP is about an order of magnitude larger than that from the local analysis, since the instability is then governed by global rather than local dynamics (see \autoref{fig10}(b)).

Finally, we compute the marginal stability curve related to the local model, i.e.~the critical Rayleigh number as a function of the aspect ratio.
The results are plotted in \autoref{fig11} and show thresholds lower than those for the global approach.
This difference tends to decrease as the aspect ratio decreases but a slight difference persists for very low aspect ratio, e.g.~$Rac=0.34$ and 0.5 at $\epsilon=1/1024$, $f=0.9$ for the local and global analysis respectively.
It should be noted, however,  that the same universal $\epsilon f^{-1}$ power law  behaviour is identified (see insert from  \autoref{fig11}).

This analysis emphasizes that such a base state, resulting from a bottom/top flux mismatch leading to vertical shear combined with horizontal and vertical temperature gradients, significantly lowers the stability threshold compared to the purely thermal threshold of Chapman \& Proctor.
The threshold value for purely thermal instability at $f=0.9$ is 640, and it remains independent of the aspect ratio.
In contrast, the threshold found from local analysis under the same conditions decreases as the aspect ratio decreases, following a $\epsilon f^{-1}$ power law and reaching a value of 0.35 for $\epsilon=1/1024$.
Similar behaviour was identified for the low-$Pr$ regime by \cite{ORTIZPEREZ2014,ORTIZPEREZ2015,Patne_Oron_2022} in the context of an inclined thermal gradient where both horizontal and vertical gradients of temperature are fixed.
In our system, the horizontal temperature gradient is a dynamic consequence of the difference in flux between the top and bottom.

The local analysis of the reduced 1D model identifies a 3D oscillatory instability in the low $\Ra$ regime.
Despite the limitations of the local approach, in particular accounting for the fact that the most unstable mode has $k_x \simeq 0$, it exhibits strong similarities to the global instability.
In the asymptotic limit of small aspect ratios, the transverse wavenumber and angular frequency converge to similar constant values, with perturbation fields displaying a similar vertical mode structure.
The dependence on the Prandtl number is also consistent between the two approaches, showing a stabilizing effect as $Pr$ increases and the disappearance of the most unstable mode around $Pr \simeq 0.3$.
Furthermore, both local and global analyses reveal the same asymptotic behavior for small aspect ratios, where the critical Rayleigh number follows a $\epsilon f^{-1}$ power law.
These shared characteristics confirm the relevance of the local model in capturing key features of the global instability.
Given its ability to reproduce many aspects of the global instability, we now leverage the analytical results of the local analysis to further investigate the underlying instability mechanisms.

\subsubsection{Instability mechanisms}
In this section, we analyze the linear perturbation equations \eqref{ucart}--\eqref{Mcart} given in Appendix~\ref{EVPloc}, further assuming a Fourier decomposition in the $z$-direction.
Performing the necessary algebraic manipulations, we obtain the following cubic dispersion relation
\begin{equation}
    \begin{aligned}
        \lambda^3 - \lambda^2(\mathcal{S}-\mathcal{D}) + \lambda( \mathcal{C} -\mathcal{S}\mathcal{D})  - \mathcal{B}=0,
    \end{aligned} 
    \label{disp}
\end{equation}
where
\begin{equation}
    \begin{aligned}
         &\mathcal{S}=U_zq_xq_z,\qquad  ~~~~~~\mathcal{C}=\Ra Pr\big[T_z(1-q_z^2)- T_xq_xq_z\big], \\ &\mathcal{D}=k^2(1-Pr),\qquad \mathcal{B}=\Ra Pr U_zT_xq_y^2,
    \end{aligned} 
    \label{termsdisp}
\end{equation}
and $\lambda = \sigma+Prk^2 + i(Uk_x+\omega )$ with $q_j = k_j/k$ for $j=x$, $y$, $z$ and $k^2=k_x^2+k_y^2+k_z^2$.
Although the Fourier decomposition in the $z$-direction is an approximation due to the top/bottom boundary conditions, this local analysis remains valuable for identifying the dominant instability mechanisms.
The dispersion relation \eqref{disp} reveals four distinct contributions:
\begin{enumerate}
    \item A shear term $\mathcal{S}$, proportional to the vertical velocity gradient $U_z$, representing a potential source for the     shear instability.
          This term contributes to 2D disturbances in the $(x,z)$ plane.
    \item A diffusive term $\mathcal{D}$, arising from the disparity between viscous and thermal diffusion.
          This contribution is spatially uniform and changes sign depending on whether the Prandtl number is greater or less than one. 
    \item A convective term $\mathcal{C}$, that includes the effects of the background temperature gradients and composed of two parts.
The first part reflects the influence of the vertical temperature gradient in the base state: it is stabilizing when $T_z > 0$ (i.e.~stably stratified) and destabilizing when $T_z < 0$, as in the classical Rayleigh--B\'enard configuration.
Based on the base state shown in \autoref{fig12}, this term is destabilizing near the center and bottom of the domain, where $T_z<0$ due to the imposed heat flux at the lower boundary.
In contrast, further from the center and in the bulk of the domain, the term becomes stabilizing ($T_z > 0$), due to re-stratification induced by horizontal advective heat transport.
This effect is captured by the term$-\epsilon^{-1}\Rar \mathrm{P}(z)G^2(x)$ in equation \eqref{tzglobal} where $\mathrm{P}(z) < 0$ for all $z \in [0,1]$.
Such restratification has also been observed by \cite{Patne_Oron_2022,Dixit_Bukhari_Patne_2024} in configurations with imposed oblique temperature gradients.

The second part of the convective term $\mathcal{C}$ involves the horizontal temperature gradient $T_x$ which is negative throughout the domain. This contribution is stabilizing for perturbations with $q_x,q_z>0$, and does not lead to disturbances in the $y$-direction.

\item  A baroclinic term $\mathcal{B}$, representing the interaction between vertical shear and horizontal temperature gradient.
It is proportional to the product $U_zT_x$, and its effect depends on the sign of this product (i.e.~the sign of $U_z$ since $T_x<0$ throughout the domain).
Moreover, this is the only term that exclusively affects transverse modes ($k_y \neq 0$).
\end{enumerate}
\begin{figure}
   \centering
    \includegraphics[scale=0.35]{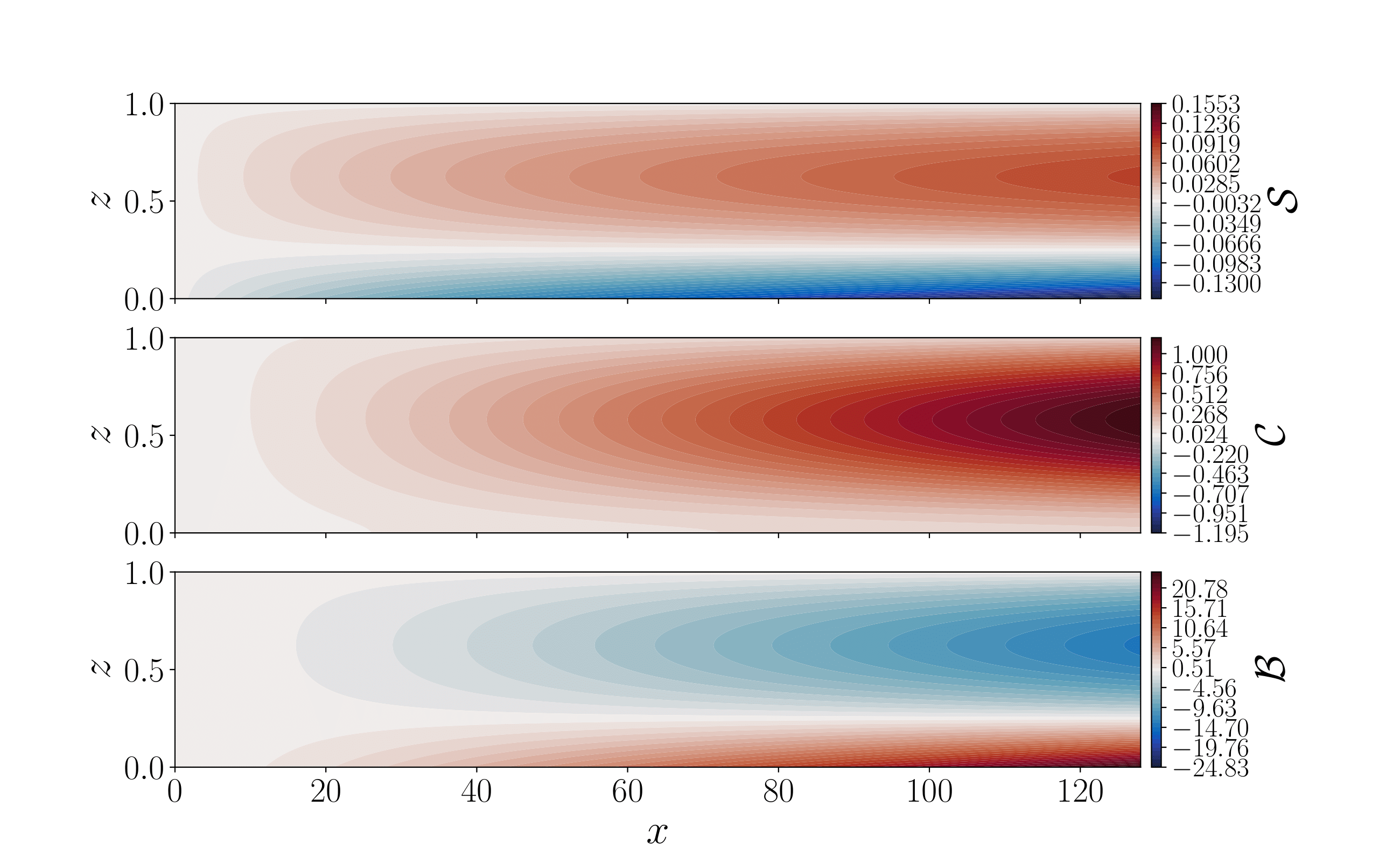}
   \caption{Maps in the $(x,z)$ plane of the shear term $\mathcal{S}$ (top), the convective term $\mathcal{C}$ (middle) and the baroclinic term $\mathcal{B}$ (bottom) for the case $\epsilon=1/128$, $f=0.9$, $Pr=0.1$ and $\Ra=2.75$, corresponding to the threshold value identified from the local EVP analysis. $k_x=1.93\times10^{-2}$, $k_y=0.70$, corresponding to the most unstable mode from the local EVP while $k_z=\pi$. The diffusive term is uniform across the domain with a constant value $\mathcal{D}=9.32$.}
   \label{fig15}
\end{figure}

The key elements associated with this mechanism are $U_z$ and $T_x$.
This coupling bears some similarities to baroclinic instability, which also relies on the interaction between vertical shear and horizontal temperature gradient.
Classical baroclinic instability, however, is derived in rotating systems, where the base flow is in thermal wind and geostrophic balance \cite[e.g.][]{pedlosky2013geophysical}.
This fundamental aspect sets baroclinic instability apart from our case.
In our configuration, there is no background rotation and  no geostrophic balance.
While our base state does feature vertical shear and horizontal temperature gradient, the nature of their interaction differs from that of baroclinic instability: the vertical shear $U_z$ is aligned with the horizontal temperature gradient $T_x$, whereas in baroclinic instability  the vertical shear is perpendicular to the thermal gradient.

\autoref{fig15} shows maps of $\mathcal{S}$, $\mathcal{C}$, and $\mathcal{B}$ for an unstable case identified via the local EVP analysis, slightly above the onset.
Here $k_x$ and $k_y$ are the wavenumbers of the most unstable local mode, while $k_z=\pi$.
The diffusive term $\mathcal{D}=9.32$ is uniform throughout the domain.
Slightly above onset, the relative magnitudes are $\|\mathcal{B}\| \gg \|\mathcal{D}\| \gg \| \mathcal{C}\| \gg\|\mathcal{S}\|$, suggesting a dominant balance $\lambda^3\approx\mathcal{B}$ from the dispersion relation \eqref{disp}.

Since the instability is oscillatory, it is expected to arise in regions where $\mathcal{B} < 0$, typically near the side in the upper half of the domain (see the bottom panel of Figure~\ref{fig13}).
In this area, the dispersion relation yields a pair of complex-conjugate eigenvalues with positive real parts. Consequently, the approximate real part of the growth rate is
\begin{equation}
    \sigma \approx |\Ra Pr U_z T_xq_y^2|^{1/3} - Prk^2.
    \label{threshold}
\end{equation}
To estimate the critical Rayleigh number at the onset ($\sigma =0$), we use scalings derived from the bulk base-state solution in \S\ref{sec:baseflow}, valid in the asymptotic regime
$\epsilon \ll 1$ with $\Ra \sim O(1)$.
In this regime, the vertical shear and horizontal temperature gradient scale as
\begin{equation}
    U_z  \sim \hat{R}a \mathrm{P}^\prime{\Theta_0}_X \sim \Ra f \epsilon^{-1}\qquad \mbox{and}\qquad  T_x \sim \epsilon^{-2}{\Theta_0}_x\sim f \epsilon^{-1}.
\label{scaling}
\end{equation}
Substituting relation \eqref{scaling} into \eqref{threshold} gives the scaling
\begin{equation}
    Ra_c \approx \frac{\epsilon Pr}{f}.
    \label{Rac_scal}
\end{equation}
This scaling is consistent with the trend observed in \autoref{fig11}, confirming that $Ra_c \sim \epsilon f^{-1}$.
Equation \eqref{Rac_scal} further indicates that the larger $Pr$ the more stable the system, as observed in both global and local analyses. 
Physically, the stabilizing effect of the Prandtl number can be interpreted as arising from viscous damping of the perturbation and from the decrease of the vertical shear amplitude, which reduces the destabilizing forcing.
Altogether, these findings suggest that the instability is driven by a baroclinic-type mechanism, resulting from the interaction between the vertical shear and the horizontal temperature gradient.

However, our analysis has not accounted for the effects of the domain edge.
It is plausible that the localized recirculation near the side could stabilize the instability described here, as this recirculation alters the flow profile and the no-slip condition at the side results in increased viscous dissipation. 
This phenomenon may explain why the critical Rayleigh number is smaller in the local approach than in the global one.
Nevertheless, we cannot rule out the possibility that this recirculation is also contributing to the observed instability.
Streamline curvature, which increases with lower $\epsilon$, combined with the horizontal temperature gradient, may induce a thermo-centrifugal type 3D instability.
\cite{Mutabazi2002,Mutabazi2013,Mutabazi2015} studied such coupling extensively in the context of a canonical Taylor-Couette flow subjected to a fixed radial temperature gradient.
Further analysis is needed to fully understand this additional effect in the global system. which we leave for future studies.

\section{Conclusion}
In this study, we analysed the flow structure for small aspect ratios of a thin fluid layer with imposed heat flux at the top and bottom and imposed temperature at the side. 
Near the edge, a vortex recirculation zone was identified, whose characteristic size is independent of the aspect ratio. 
We described the bulk base state using an asymptotic approach based on lubrication theory, which accurately captures the vertical and horizontal structures of the velocity and temperature fields in the small aspect ratio regime.
The bulk solution shows some expected differences only in the vortex region near the sidewall.

\begin{figure}
   \centering
    \includegraphics[scale=0.32]{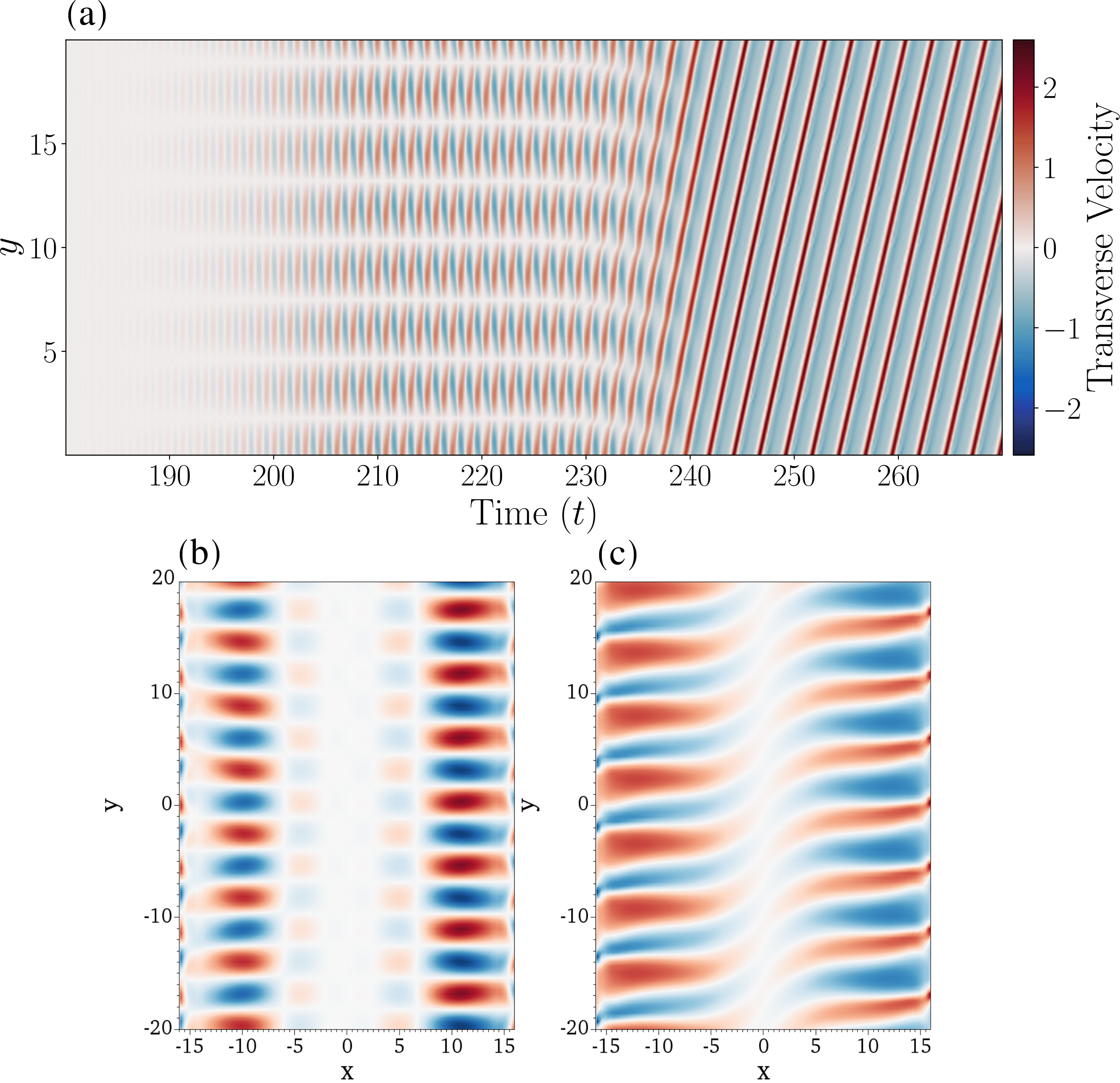}
   \caption{$(a)$ Hovm\"oller(space-time) diagram of the $z$-averaged transverse velocity measured at $X=3/4$. 2D snapshots in the $(x,y)$-plan of the transverse velocity at $(b)$ $t = 210$ and $(c)$ $t = 260$. The IVP-DNS method is used to obtain the data which correspond to the simulation IVP11 listed in \autoref{table:1}.  The parameters are $\epsilon=1/16$,$L_y=40$, $Pr=0.1$, $f=0.9$ and $\Ra=170$. }
   \label{fig16}
\end{figure}

A detailed characterization of system instability revealed that the most unstable mode is three-dimensional, featuring two superimposed oscillatory standing waves in the transverse direction close to the domain edge.
During the exponential growth phase of the instability, these waves interact non-linearly, ultimately producing a drift pattern as illustrated in \autoref{fig16}.
A preferred drift direction is selected, either forward or backward (clockwise or counterclockwise in cylindrical geometry), consistent with the findings of \cite{reinjfm}.
A weakly non-linear study is necessary in order to understand the mechanism at the origin of the drift.
The instability threshold decreases with the normalized flux difference and increases with aspect ratio, following a $f^{-1}\epsilon$ power law behaviour.
We found that increasing the Prandtl number stabilizes the system and spreads out the mode structure localised near the edge to the bulk without changing the transverse wavenumber.
For $Pr\simeq0.3$ a complete change in modal structure is identified, the transverse wavenumber changes and the mode structure becomes localised around the center.

We proposed a 1D reduced model based on the analytical solution for the bulk base state in order to simplify the system and capture the physical mechanisms responsible for the global instability.
We performed a local stability analysis of this model and showed oscillating transverse instabilities for unexpectedly low Rayleigh thresholds.
However, we also found instability for horizontal wavelengths similar to the domain size, challenging our local analysis assumption.

Despite the limitations of the local approach it exhibits strong similarities with the global instability.
In the asymptotic limit of small aspect ratios, the transverse wavenumber and angular frequency converge to similar constant values, with perturbation fields displaying a similar vertical mode structure.
The dependence on the Prandtl number is also consistent between the two approaches, showing a stabilizing effect as $Pr$ increases and the disappearance of the mode around $Pr \simeq 0.3$.
Furthermore, both local and global analyses reveal the same asymptotic behavior for small aspect ratios, where the critical Rayleigh number follows a $\epsilon f^{-1}$ power law.
These shared characteristics confirm the relevance of the local model in capturing key features of the global instability.

In our local analysis, we identified three mechanisms contributing to instability: a vertical shear, a Rayleigh-B\'enard-type thermal forcing, and a third mechanism arising from the coupling between vertical shear and horizontal temperature gradient, reminiscent of baroclinic-type instabilities.
By solving the perturbation equations in Fourier modes, we showed that this coupling produces the same $\epsilon f^{-1}$ scaling law for the critical Rayleigh number and it also captures the stabilizing effect of increasing the Prandtl number.
Although all mechanisms give rise to identifiable modes in the stability analysis, the baroclinic-type mechanism is the most unstable with surprisingly low critical Rayleigh number threshold, and is thus most likely responsible for the dominant unstable mode observed at low Prandtl numbers in the DNS.

To return to the nuclear context motivating this study, the patterns observed here persist into the turbulent regime \citep{reinjfm}. \citet{reinmanip} showed experimentally in cylindrical geometry that drifting patterns are responsible for transporting a significant fraction of the heat flux along the sidewalls, thereby contributing to localized abrasion of the vessel wall.  
The present analysis provides new insight into the physical mechanisms underlying the patterns' emergence, which may inform future supression or control strategies.  
An interesting implication of these results is that modifying the composition of the metallic layer, and therefore its effective Prandtl number, could alter the structure of the unstable modes and consequently modify the drifting pattern and sidewall transport.
While this points to a possible route for mitigating wall-localized fluxes, any practical application would be subject to significant material, structural and safety constraints, and lies beyond the scope of the present study.

\section*{Acknowledgements}
This work was supported by the Institut de Radioprotection et de Sûreté Nucléaire (IRSN), the Commissariat à l'Énergie Atomique et aux Énergies Alternatives (CEA), Électricité de France (EDF), and the Simons Foundation. 
Centre de Calcul Intensif d’Aix-Marseille is acknowledged for granting access to its high performance computing resources.
This work was granted access to the HPC resources of IDRIS under the allocation 2025-A0180407543 made by GENCI.
FR, BF and MLB would like to thank Florian Fichot, Laure Carenini and Jean-Marc Ricaud from the Nuclear Safety and Radiation Protection Authority for bringing this problem to their attention and for further fruitful discussions.

\section*{Declaration of Interests}
The authors declare no competing interests. The funding organizations—IRSN, CEA, EDF, and the Simons Foundation—had no involvement in the study design, data collection, analysis, interpretation, or the decision to publish.
Responsibility for the content lies solely with the authors.

\appendix
\section{Summary of the simulation parameters}\label{A}
Table \ref{table:1} provide all relevant parameters for the simulations performed using the IVP, BVP, Global-EVP and Local-EVP approaches.
\begin{table}
\centering
\begin{tabular}{lllllllll}
\hline
\multicolumn{1}{c}{Name} & \multicolumn{1}{c}{Method} & \multicolumn{1}{c}{$\Ra$} & \multicolumn{1}{c}{$\epsilon$} & \multicolumn{1}{c}{$f$} & \multicolumn{1}{c}{$Pr$} & \multicolumn{1}{c}{$k_x$} & \multicolumn{1}{c}{$k_y$} & \multicolumn{1}{c}{$N$} \\ \hline
\multicolumn{1}{c}{IVP1} & \multicolumn{1}{c}{IVP} & \multicolumn{1}{c}{10} & \multicolumn{1}{c}{1/8} & \multicolumn{1}{c}{0.9} & \multicolumn{1}{c}{0.1} & \multicolumn{1}{c}{-} & \multicolumn{1}{c}{-} & \multicolumn{1}{c}{6} \\
\multicolumn{1}{c}{IVP2} & \multicolumn{1}{c}{IVP} & \multicolumn{1}{c}{10} & \multicolumn{1}{c}{1/16} & \multicolumn{1}{c}{0.9} & \multicolumn{1}{c}{0.1} & \multicolumn{1}{c}{-} & \multicolumn{1}{c}{-} & \multicolumn{1}{c}{6} \\
\multicolumn{1}{c}{IVP3} & \multicolumn{1}{c}{IVP} & \multicolumn{1}{c}{10} & \multicolumn{1}{c}{1/32} & \multicolumn{1}{c}{0.9} & \multicolumn{1}{c}{0.1} & \multicolumn{1}{c}{-} & \multicolumn{1}{c}{-} & \multicolumn{1}{c}{6} \\
\multicolumn{1}{c}{BVP1} & \multicolumn{1}{c}{BVP} & \multicolumn{1}{c}{$\{0.98;988\}_{20}$} & \multicolumn{1}{c}{1/32} & \multicolumn{1}{c}{0.9} & \multicolumn{1}{c}{0.1} & \multicolumn{1}{c}{-} & \multicolumn{1}{c}{-} & \multicolumn{1}{c}{16} \\
\multicolumn{1}{c}{BVP2} & \multicolumn{1}{c}{BVP} & \multicolumn{1}{c}{$\{0.49;494\}_{20}$} & \multicolumn{1}{c}{1/64} & \multicolumn{1}{c}{0.9} & \multicolumn{1}{c}{0.1} & \multicolumn{1}{c}{-} & \multicolumn{1}{c}{-} & \multicolumn{1}{c}{16} \\
\multicolumn{1}{c}{BVP3} & \multicolumn{1}{c}{BVP} & \multicolumn{1}{c}{$\{0.24;247\}_{20}$} & \multicolumn{1}{c}{1/128} & \multicolumn{1}{c}{0.9} & \multicolumn{1}{c}{0.1} & \multicolumn{1}{c}{-} & \multicolumn{1}{c}{-} & \multicolumn{1}{c}{18} \\
\multicolumn{1}{c}{BVP4} & \multicolumn{1}{c}{BVP} & \multicolumn{1}{c}{$\{0.12;123\}_{20}$} & \multicolumn{1}{c}{1/256} & \multicolumn{1}{c}{0.9} & \multicolumn{1}{c}{0.1} & \multicolumn{1}{c}{-} & \multicolumn{1}{c}{-} & \multicolumn{1}{c}{18} \\
\multicolumn{1}{c}{BVP5} & \multicolumn{1}{c}{BVP} & \multicolumn{1}{c}{$\{0.06;61\}_{20}$} & \multicolumn{1}{c}{1/512} & \multicolumn{1}{c}{0.9} & \multicolumn{1}{c}{0.1} & \multicolumn{1}{c}{-} & \multicolumn{1}{c}{-} & \multicolumn{1}{c}{20} \\
\multicolumn{1}{c}{BVP6} & \multicolumn{1}{c}{BVP} & \multicolumn{1}{c}{$\{0.03;31\}_{20}$} & \multicolumn{1}{c}{1/1024} & \multicolumn{1}{c}{0.9} & \multicolumn{1}{c}{0.1} & \multicolumn{1}{c}{-} & \multicolumn{1}{c}{-} & \multicolumn{1}{c}{20} \\
\multicolumn{1}{c}{IVP4} & \multicolumn{1}{c}{IVP} & \multicolumn{1}{c}{1} & \multicolumn{1}{c}{1/32} & \multicolumn{1}{c}{0.9} & \multicolumn{1}{c}{0.1} & \multicolumn{1}{c}{-} & \multicolumn{1}{c}{-} & \multicolumn{1}{c}{6} \\
\multicolumn{1}{c}{IVP5} & \multicolumn{1}{c}{IVP} & \multicolumn{1}{c}{3} & \multicolumn{1}{c}{1/32} & \multicolumn{1}{c}{0.9} & \multicolumn{1}{c}{0.1} & \multicolumn{1}{c}{-} & \multicolumn{1}{c}{-} & \multicolumn{1}{c}{6} \\
\multicolumn{1}{c}{IVP6} & \multicolumn{1}{c}{IVP} & \multicolumn{1}{c}{10} & \multicolumn{1}{c}{1/32} & \multicolumn{1}{c}{0.9} & \multicolumn{1}{c}{0.1} & \multicolumn{1}{c}{-} & \multicolumn{1}{c}{-} & \multicolumn{1}{c}{6} \\
\multicolumn{1}{c}{IVP7} & \multicolumn{1}{c}{IVP} & \multicolumn{1}{c}{30} & \multicolumn{1}{c}{1/32} & \multicolumn{1}{c}{0.9} & \multicolumn{1}{c}{0.1} & \multicolumn{1}{c}{-} & \multicolumn{1}{c}{-} & \multicolumn{1}{c}{6} \\
\multicolumn{1}{c}{IVP8} & \multicolumn{1}{c}{IVP} & \multicolumn{1}{c}{100} & \multicolumn{1}{c}{1/32} & \multicolumn{1}{c}{0.9} & \multicolumn{1}{c}{0.1} & \multicolumn{1}{c}{-} & \multicolumn{1}{c}{-} & \multicolumn{1}{c}{8} \\
\multicolumn{1}{c}{IVP9} & \multicolumn{1}{c}{IVP} & \multicolumn{1}{c}{300} & \multicolumn{1}{c}{1/32} & \multicolumn{1}{c}{0.9} & \multicolumn{1}{c}{0.1} & \multicolumn{1}{c}{-} & \multicolumn{1}{c}{-} & \multicolumn{1}{c}{8} \\
\multicolumn{1}{c}{IVP10} & \multicolumn{1}{c}{IVP} & \multicolumn{1}{c}{100} & \multicolumn{1}{c}{1/512} & \multicolumn{1}{c}{0.9} & \multicolumn{1}{c}{0.1} & \multicolumn{1}{c}{-} & \multicolumn{1}{c}{-} & \multicolumn{1}{c}{8} \\
\multicolumn{1}{c}{IVP11} & \multicolumn{1}{c}{IVP} & \multicolumn{1}{c}{170} & \multicolumn{1}{c}{1/16} & \multicolumn{1}{c}{0.9} & \multicolumn{1}{c}{0.1} & \multicolumn{1}{c}{-} & \multicolumn{1}{c}{-} & \multicolumn{1}{c}{6} \\
\multicolumn{1}{c}{IVP12} & \multicolumn{1}{c}{IVP} & \multicolumn{1}{c}{5000} & \multicolumn{1}{c}{1/16} & \multicolumn{1}{c}{0.9} & \multicolumn{1}{c}{1.0} & \multicolumn{1}{c}{-} & \multicolumn{1}{c}{-} & \multicolumn{1}{c}{8} \\
\multicolumn{1}{c}{GEVP1} & \multicolumn{1}{c}{Global-EVP} & \multicolumn{1}{c}{700} & \multicolumn{1}{c}{1/8} & \multicolumn{1}{c}{[0.05;0.95]} & \multicolumn{1}{c}{0.1} & \multicolumn{1}{c}{-} & \multicolumn{1}{c}{$[0.01;50]_{50}$} & \multicolumn{1}{c}{16} \\
\multicolumn{1}{c}{GEVP2} & \multicolumn{1}{c}{Global-EVP} & \multicolumn{1}{c}{150} & \multicolumn{1}{c}{1/16} & \multicolumn{1}{c}{[0.05;0.9]} & \multicolumn{1}{c}{0.1} & \multicolumn{1}{c}{-} & \multicolumn{1}{c}{$[0.01;50]_{50}$} & \multicolumn{1}{c}{16} \\
\multicolumn{1}{c}{GEVP3} & \multicolumn{1}{c}{Global-EVP} & \multicolumn{1}{c}{31} & \multicolumn{1}{c}{1/32} & \multicolumn{1}{c}{[0.05;0.9]} & \multicolumn{1}{c}{0.1} & \multicolumn{1}{c}{-} & \multicolumn{1}{c}{$[0.01;50]_{50}$} & \multicolumn{1}{c}{16} \\
\multicolumn{1}{c}{GEVP4} & \multicolumn{1}{c}{Global-EVP} & \multicolumn{1}{c}{12} & \multicolumn{1}{c}{1/64} & \multicolumn{1}{c}{[0.05;0.9]} & \multicolumn{1}{c}{0.1} & \multicolumn{1}{c}{-} & \multicolumn{1}{c}{$[0.01;50]_{50}$} & \multicolumn{1}{c}{18} \\
\multicolumn{1}{c}{GEVP5} & \multicolumn{1}{c}{Global-EVP} & \multicolumn{1}{c}{5} & \multicolumn{1}{c}{1/128} & \multicolumn{1}{c}{[0.05;0.9]} & \multicolumn{1}{c}{0.1} & \multicolumn{1}{c}{-} & \multicolumn{1}{c}{$[0.01;50]_{50}$} & \multicolumn{1}{c}{18} \\
\multicolumn{1}{c}{GEVP6} & \multicolumn{1}{c}{Global-EVP} & \multicolumn{1}{c}{2.1} & \multicolumn{1}{c}{1/256} & \multicolumn{1}{c}{[0.05;0.9]} & \multicolumn{1}{c}{0.1} & \multicolumn{1}{c}{-} & \multicolumn{1}{c}{$[0.6;0.8,1.5]$} & \multicolumn{1}{c}{18} \\
\multicolumn{1}{c}{GEVP7} & \multicolumn{1}{c}{Global-EVP} & \multicolumn{1}{c}{1} & \multicolumn{1}{c}{1/512} & \multicolumn{1}{c}{[0.05;0.9]} & \multicolumn{1}{c}{0.1} & \multicolumn{1}{c}{-} & \multicolumn{1}{c}{$[0.6;0.8,1.5]$} & \multicolumn{1}{c}{20} \\
\multicolumn{1}{c}{GEVP8} & \multicolumn{1}{c}{Global-EVP} & \multicolumn{1}{c}{0.5} & \multicolumn{1}{c}{1/1024} & \multicolumn{1}{c}{[0.05;0.9]} & \multicolumn{1}{c}{0.1} & \multicolumn{1}{c}{-} & \multicolumn{1}{c}{$[0.6;0.8,1.5]$} & \multicolumn{1}{c}{20} \\
\multicolumn{1}{c}{GEVP9} & \multicolumn{1}{c}{Global-EVP} & \multicolumn{1}{c}{$\{3.1,980\}_{11}$} & \multicolumn{1}{c}{1/32} & \multicolumn{1}{c}{0.9} & \multicolumn{1}{c}{0.1} & \multicolumn{1}{c}{-} & \multicolumn{1}{c}{$[0.01;50]_{50}$} & \multicolumn{1}{c}{16} \\
\multicolumn{1}{c}{GEVP10} & \multicolumn{1}{c}{Global-EVP} & \multicolumn{1}{c}{$\{150;15000\}_{11}$} & \multicolumn{1}{c}{1/16} & \multicolumn{1}{c}{0.9} & \multicolumn{1}{c}{0.13} & \multicolumn{1}{c}{-} & \multicolumn{1}{c}{$[0.6;2]_{5}$} & \multicolumn{1}{c}{20} \\
\multicolumn{1}{c}{GEVP11} & \multicolumn{1}{c}{Global-EVP} & \multicolumn{1}{c}{$\{150;15000\}_{11}$} & \multicolumn{1}{c}{1/16} & \multicolumn{1}{c}{0.9} & \multicolumn{1}{c}{0.16} & \multicolumn{1}{c}{-} & \multicolumn{1}{c}{$[0.6;2]_{5}$} & \multicolumn{1}{c}{20} \\
\multicolumn{1}{c}{GEVP12} & \multicolumn{1}{c}{Global-EVP} & \multicolumn{1}{c}{$\{150;15000\}_{11}$} & \multicolumn{1}{c}{1/16} & \multicolumn{1}{c}{0.9} & \multicolumn{1}{c}{0.2} & \multicolumn{1}{c}{-} & \multicolumn{1}{c}{$[0.6;2]_{5}$} & \multicolumn{1}{c}{20} \\
\multicolumn{1}{c}{GEVP13} & \multicolumn{1}{c}{Global-EVP} & \multicolumn{1}{c}{$\{150;15000\}_{11}$} & \multicolumn{1}{c}{1/16} & \multicolumn{1}{c}{0.9} & \multicolumn{1}{c}{0.23} & \multicolumn{1}{c}{-} & \multicolumn{1}{c}{$[0.6;2]_{5}$} & \multicolumn{1}{c}{20} \\
\multicolumn{1}{c}{GEVP14} & \multicolumn{1}{c}{Global-EVP} & \multicolumn{1}{c}{$\{150;15000\}_{11}$} & \multicolumn{1}{c}{1/16} & \multicolumn{1}{c}{0.9} & \multicolumn{1}{c}{0.26} & \multicolumn{1}{c}{-} & \multicolumn{1}{c}{$[0.6;2]_{5}$} & \multicolumn{1}{c}{20} \\
\multicolumn{1}{c}{GEVP15} & \multicolumn{1}{c}{Global-EVP} & \multicolumn{1}{c}{$\{150;15000\}_{11}$} & \multicolumn{1}{c}{1/16} & \multicolumn{1}{c}{0.9} & \multicolumn{1}{c}{0.3} & \multicolumn{1}{c}{-} & \multicolumn{1}{c}{$[0.6;2]_{5}$} & \multicolumn{1}{c}{20} \\
\multicolumn{1}{c}{GEVP16} & \multicolumn{1}{c}{Global-EVP} & \multicolumn{1}{c}{$\{10^{4};10^6\}_{11}$} & \multicolumn{1}{c}{1/16} & \multicolumn{1}{c}{0.9} & \multicolumn{1}{c}{0.4} & \multicolumn{1}{c}{-} & \multicolumn{1}{c}{$[0.6;2]_{5}$} & \multicolumn{1}{c}{20} \\
\multicolumn{1}{c}{LEVP1} & \multicolumn{1}{c}{Local-EVP} & \multicolumn{1}{c}{$21.27$} & \multicolumn{1}{c}{1/16} & \multicolumn{1}{c}{0.9} & \multicolumn{1}{c}{0.1} & \multicolumn{1}{c}{$[-2;2]_{48}$} & \multicolumn{1}{c}{$[0;2]_{30}$} & \multicolumn{1}{c}{128} \\
\multicolumn{1}{c}{LEVP2} & \multicolumn{1}{c}{Local-EVP} & \multicolumn{1}{c}{$\{1.37,10.84\}_{4}$} & \multicolumn{1}{c}{$\{1/32;1/256\}_{4}$} & \multicolumn{1}{c}{0.9} & \multicolumn{1}{c}{0.1} & \multicolumn{1}{c}{$[-2;2]_{48}$} & \multicolumn{1}{c}{$[0;2]_{30}$} & \multicolumn{1}{c}{128} \\
\multicolumn{1}{c}{LEVP3} & \multicolumn{1}{c}{Local-EVP} & \multicolumn{1}{c}{$3.66$} & \multicolumn{1}{c}{$1/96$} & \multicolumn{1}{c}{0.9} & \multicolumn{1}{c}{0.1} & \multicolumn{1}{c}{$[-2;2]_{48}$} & \multicolumn{1}{c}{$[0;2]_{30}$} & \multicolumn{1}{c}{128} \\
\multicolumn{1}{c}{LEVP4} & \multicolumn{1}{c}{Local-EVP} & \multicolumn{1}{c}{$1.10$} & \multicolumn{1}{c}{$1/320$} & \multicolumn{1}{c}{0.9} & \multicolumn{1}{c}{0.1} & \multicolumn{1}{c}{$[-2;2]_{48}$} & \multicolumn{1}{c}{$[0;2]_{30}$} & \multicolumn{1}{c}{128} \\
\multicolumn{1}{c}{LEVP5} & \multicolumn{1}{c}{Local-EVP} & \multicolumn{1}{c}{$10.84$} & \multicolumn{1}{c}{$1/32$} & \multicolumn{1}{c}{0.9} & \multicolumn{1}{c}{0.1} & \multicolumn{1}{c}{$[-2;2]_{48}$} & \multicolumn{1}{c}{$[0;2]_{30}$} & \multicolumn{1}{c}{128} \\
\multicolumn{1}{c}{LEVP6} & \multicolumn{1}{c}{Local-EVP} & \multicolumn{1}{c}{$5.46$} & \multicolumn{1}{c}{$1/64$} & \multicolumn{1}{c}{0.9} & \multicolumn{1}{c}{0.1} & \multicolumn{1}{c}{$[-2;2]_{48}$} & \multicolumn{1}{c}{$[0;2]_{30}$} & \multicolumn{1}{c}{128} \\
\multicolumn{1}{c}{LEVP7} & \multicolumn{1}{c}{Local-EVP} & \multicolumn{1}{c}{$2.75$} & \multicolumn{1}{c}{$1/128$} & \multicolumn{1}{c}{0.9} & \multicolumn{1}{c}{0.1} & \multicolumn{1}{c}{$[-2;2]_{48}$} & \multicolumn{1}{c}{$[0;2]_{30}$} & \multicolumn{1}{c}{128} \\
\multicolumn{1}{c}{LEVP8} & \multicolumn{1}{c}{Local-EVP} & \multicolumn{1}{c}{$0.68$} & \multicolumn{1}{c}{$1/512$} & \multicolumn{1}{c}{0.9} & \multicolumn{1}{c}{0.1} & \multicolumn{1}{c}{$[-2;2]_{48}$} & \multicolumn{1}{c}{$[0;2]_{30}$} & \multicolumn{1}{c}{128} \\
\multicolumn{1}{c}{LEVP9} & \multicolumn{1}{c}{Local-EVP} & \multicolumn{1}{c}{$0.34$} & \multicolumn{1}{c}{$1/1024$} & \multicolumn{1}{c}{0.9} & \multicolumn{1}{c}{0.1} & \multicolumn{1}{c}{$[-2;2]_{48}$} & \multicolumn{1}{c}{$[0;2]_{30}$} & \multicolumn{1}{c}{128} \\
\hline
\end{tabular}
\caption{List of simulations (DNS IVP/BVP and EVP Global/Local) with physical and numerical parameters. $N$ is the order of the spectral modes (same for each direction). The notation $[A,B]_{n}$ denotes an array of $n$ linearly spaced values from $A$ to $B$, inclusive, while $\{A,B\}_{n}$ represents an array of $n$ values spaced evenly on a logarithmic scale from $A$ to $B$ inclusive.  } 
\label{table:1}
\end{table}

\section{Numerical methods}\label{appendixB}
\subsection{Eigenvalue analysis}\label{PERTEQ}
To carry out an eigenvalue analysis, we consider the background state, invariant in the $y$-direction, with a pressure and temperature profile denoted $P(x,z)$ and $T(x,z)$ and with a 2D base flow $\bm{\mathrm{U}}=\left(U(x,z),0,W(x,z)\right)$. 
This base state can come from analytical considerations or from solving BVP, as shown below. 
We then consider perturbations to the base state with pressure and temperature perturbations denoted $p'(x,y,z)$ and $\theta(x,y,z)$ and with a perturbation flow $\bm{u'}=\left(u'(x,y,z),v'(x,y,z),w'(x,y,z)\right)$.

The linearized conservation equations of momentum, mass and energy become 
\begin{equation}
u'_t +\left(\bm{\mathrm{U} \cdot \nabla} \right)u' + \left(\bm{u' \cdot \nabla} \right)U=-p'_{x} + Pr \nabla^2u',
\label{qu}
\end{equation}
\begin{equation}
v'_{t} +\left(\bm{\mathrm{U} \cdot \nabla} \right)v'=-p'_{y} + Pr \nabla^2v',
\label{qv}
\end{equation}
\begin{equation}
w'_{t}+\left(\bm{\mathrm{U} \cdot \nabla} \right)w' + \left(\bm{\mathrm{u}^\prime \cdot \nabla} \right)W=-p'_{z} + \Ra Pr \theta +Pr \nabla^2w',
\label{qw}
\end{equation}
\begin{equation}
u'_{x}+v'_{y}+w'_{z} = 0,
\label{divu}
\end{equation}
\begin{equation}
\theta_{t}+\left(\bm{\mathrm{U} \cdot \nabla} \right)\theta+\left(\bm{u' \cdot \nabla} \right)T= \nabla^2\theta,
\label{heat}
\end{equation}
with the boundary conditions
\begin{equation}
\begin{aligned}
\bm{u}'=\bm{0} \quad \mbox{and}~~\theta_z=0 \quad \mbox{at}~z=0 \ ,\\
u'_{z}=v'_{z}=w'=0 \quad \mbox{and}~~\theta_{z}=0 \quad \mbox{at}~z=1\ ,\\
\bm{u}'=\bm{0} \quad \mbox{and}~~\theta=0 \quad \mbox{at}~x=\epsilon^{-1}.
\end{aligned}
\label{bcnpertu}
\end{equation}
This problem leads to two types of eigenvalue problems depending on the base state. 
A Global Eigenvalue Problem, denoted \textit{Global EVP}, which involves a two-dimensional base state numerically computed from a BVP, where the flow and temperature fields vary in both the $x$- and  $z$-directions, and a Local Eigenvalue Problem, denoted \textit{Local-EVP}, with a base state coming from an analytical solution (detailed in \autoref{sec:anasol}) for which the system is assumed to be homogeneous, or at least slowly varying, in the  $x$-direction.
In the following subsection, each method is introduced with appropriate notation and perturbation equations.
\subsubsection{Global EVP problem}
We seek a complex solution for the perturbation fields in the form of normal modes in the $y$ direction as follows
\begin{equation}
\begin{bmatrix}
u',
v',
w',
\theta,
p'
\end{bmatrix}  =  
\begin{bmatrix}
\hat{u}(x,z),~
\hat{v}(x,z),~
\hat{w}(x,z),~
\hat{\theta}(x,z),~
\hat{p}(x,z)
\end{bmatrix}  e^{\hat{s} t + ik_y y},
\end{equation}
where $\hat{s}$ represents the complex growth rate and  $k_y$, the wavenumber in the $y$-direction. 
Substituting this into \eqref{qu}--\eqref{heat}, leads to 
\begin{align}
\hat{s}\hat{u} +U\hat{u}_x + W\hat{u}_z +\hat{u}U_x + \hat{w}U_z  &= -\hat{p}_x  + Pr\mathcal{L}\hat{u}, \label{ucartG} \\
\hat{s}\hat{v} +U\hat{v}_x + W\hat{v}_z  &=   -ik_y\hat{p} + Pr\mathcal{L}\hat{v},\label{vcartG}\\ 
\hat{s}\hat{w} +U\hat{w}_x + W\hat{w}_z +\hat{u}W_x + \hat{w}W_z  &=  -\hat{p}_{z}  +\Ra Pr\hat{\theta} + Pr\mathcal{L}\hat{w},\label{wcartG} \\ 
\hat{s}\hat{\theta} +U\hat{\theta}_x + W\hat{\theta}_z +\hat{u}T_x + \hat{w}T_z  &=  \mathcal{L}\hat{\theta} ,\label{TcartG} \\
 0&= \hat{u}_x + ik_y \hat{v} + \hat{w}_z,\label{McartG}
\end{align}
with the operator $\mathcal{L} = (\partial_x^2+\partial_z^2-k_y^2)$.
\subsubsection{Local EVP problem}\label{EVPloc}
In section \ref{LocalEVP}, we use this method with a simplified 1D base state taken at a given $x$ and neglecting the $x$-dependence. 
The vertical velocity from the base state $W$ is also neglected. 
Therefore, the base flow is $\bm{\mathrm{U}}=\left(U(z),0,0\right)$.
We also account for the horizontal and vertical temperature gradient, $T_x$ and $T_z(z)$, respectively.
We seek a complex solution for the perturbation fields in a form of normal modes in the $x$-and $y$-directions as follows
\begin{equation}
\begin{bmatrix}
u',
v',
w',
\theta,
p'
\end{bmatrix}  =  
\begin{bmatrix}
\hat{u}(z),~
\hat{v}(z),~
\hat{w}(z),~
\hat{\theta}(z),~
\hat{p}(z) 
\end{bmatrix}  e^{ st + i(k_x x+k_y y)},
\end{equation}
where $s=\sigma + i\omega$ represents the complex growth rate and $k_x$, $k_y$ are the wavenumbers in the $x$-and $y$-directions, respectively.
Substituting this in \eqref{qu}--\eqref{heat} leads to 
\begin{align}
\lambda\hat{u} &= -\hat{w}U_z -ik_x\hat{p} + Pr\hat{u}_{zz}, \label{ucart} \\
\lambda\hat{v} &=   -ik_y\hat{p} + Pr\hat{v}_{zz},\label{vcart}\\ 
\lambda\hat{w} &=  -\hat{p}_{z}  +\Ra Pr\hat{\theta} + Pr\hat{w}_{zz},\label{wcart} \\ 
\lambda\hat{\theta} &= -\hat{u}T_x -\hat{w}T_z -k^2(1-Pr)~\hat{\theta}+ \hat{\theta}_{zz},\label{Tcart} \\
 0&= ik_x \hat{u}+ ik_y \hat{v} + \hat{w}_z,\label{Mcart}
\end{align}
where $\lambda=\sigma +Prk^2 + i(\omega+Uk_x)$ and $k^2=k_x^2+k_y^2$.

\section{Generalized Chapman \& Proctor instability}\label{Gcp}
\subsection{Stability analysis}
In this section, we examine the effect of differential heating between the top and bottom boundaries, disregarding sidewall effects and the presence of a base flow. 
This setup is an extension of that studied by \cite{cp} (CP), who considered various flow upper and lower boundary conditions
with equal and opposite upper and lower fluxes.
It corresponds to our setup with $f=0$ and a free-slip upper boundary.
Note that \cite{Shivashinsky} considered the case $f=1$.

To address the heat flux mismatch in our case, we generalize CP's approach by examining stability for any normalized flux difference configuration. 
The base state is motionless and subject to a temperature profile denoted $T^b$. 
A uniform volume sink term $f$ is added to the base state temperature equation to balance the flux mismatch, leading to the steady-state balance $\nabla^2 T^b=f$.
By integrating this equation twice in the $z$-direction, we obtain the parabolic temperature profile
\begin{equation}
\begin{aligned}
T^b(z) = f(\tfrac{1}{2} z^2 - \tfrac{1}{6}) - z + \tfrac{1}{2},
\end{aligned}
\end{equation}
which is the "boundary temperature" defined in \eqref{boundaryTemp} and has zero vertical average.
In the subsequent analysis, we adopt the method used by CP.
We seek a time-dependent solution for a heat flux through the layer that is just above the threshold necessary for infinitesimal motion to occur. 
The Rayleigh number is set to $\Ra=\Racp+\epsilon^2$, with $\epsilon \ll 1$ and $\Racp$ the critical Rayleigh number.
Decompose the temperature into the base state and perturbation $T=T^b + \theta$ with $\theta$ and $\bm{\mathrm{u}}$ the temperature and velocity perturbation fields, respectively. 
The momentum, mass and energy conservation equations for these perturbations are
\begin{gather}
    Pr^{-1}\left(\bm{\mathrm{u}}_{t}+ \bm{\mathrm{u}\cdot\nabla}\bm{\mathrm{u}}\right)=- \bm{\nabla}P + (\Racp+\epsilon^2)\theta \bm{e}_z+ \bm{\nabla}^2\bm{\mathrm{u}},\label{cp_qdm}  \\
    \bm{\nabla\cdot \mathrm{u}}=0 ,  \\ 
    \theta_{t}+ \bm{\mathrm{u}\cdot\nabla}\theta + w~T^b_{z}= \nabla^2 \theta. \label{cp_nrj} 
\end{gather}
CP shown that the onset of convection is a monotonically decreasing function of the wavelength, indicating that the lowest critical value of the Rayleigh number occurs for infinitely long horizontal scales. 
Consequently, we set $\partial_{x} = \epsilon \partial_{X}$.
CP identified the relevant time scaling as $\partial_{t} = \epsilon^4 \partial_{\tau}$.
$\Ra$ is just above the threshold, resulting in slow growth rates ($\epsilon^2$), and long horizontal scales, which require $O(\epsilon^{-2})$ time to diffuse information across the domain;
$y$-invariance is assumed and the velocity field can be written in terms of the streamfunction $\bm{\mathrm{u}}=(\psi_{z},0,-\epsilon\psi_{X})$.
Adopting the scalings mentioned above and in addition setting $\psi=\epsilon\xi$, \eqref{cp_nrj} and the curl of \eqref{cp_qdm} lead to
\begin{equation}
\begin{aligned}
Pr^{-1}\left[\epsilon^6 \xi_{\tau XX} +\epsilon^4 \xi_{\tau zz} + \epsilon^4 \mathcal{J}(\xi,\xi_{XX}) +\epsilon^2 \mathcal{J}(\xi, \xi_{zz})\right]\\
=(\Racp+\epsilon^2)\theta_X + \epsilon^4 \xi_X +2\epsilon^2\xi_{XXzz} +\xi_{zzzz},
\label{cp_big}
\end{aligned}
\end{equation}
\begin{equation}
\begin{aligned}
\epsilon^4 \theta_{\tau} +\epsilon^2\mathcal{J}(\xi,\theta) =
\epsilon^2 \theta_{XX} +\theta_{zz} -\epsilon^2 \xi_{X}~T^b_{z}.
\label{cp_bigT}
\end{aligned}
\end{equation}
Since odd powers of $\epsilon$ are absent in the expansion, we expand $\theta$ and $\xi$ as power series in $\epsilon^2$, yielding
\begin{equation}
\begin{aligned}
\theta &= \theta_0(X,z,\tau) +  \epsilon^2 \theta_2(X,z,\tau) + O(\epsilon^4), \\
\xi &= \xi_0(X,z,\tau) +  \epsilon^2 \xi_2(X,z,\tau) + O(\epsilon^4).
\end{aligned}
\end{equation}
We now substitute these expansions into \eqref{cp_big} and \eqref{cp_bigT}.
At $O(1)$, the equations yield
\begin{equation}
       \theta_{zz} = 0 \qquad \mbox{and} \qquad \Racp \theta_{X} +\xi_{zzzz}= 0.
\end{equation}
Since no heat flux is assumed to be transported by the perturbations at the boundaries, we have $\theta_z=0$.
Consequently, $\theta_0= \bar{\theta_0}(X,\tau)$ leading to the streamfunction $\xi_0 = \Racp \mathrm{P}(z) {\bar{\theta_0}}_{X}$, which is the same equation as \eqref{0psi}, with $\mathrm{P}(z)$ defined in \eqref{P}.
This relation results from the same force balance between viscosity and buoyancy.
Considering the $O(\epsilon^2)$ terms of the temperature equation yields
\begin{equation}
\begin{aligned}
\mathcal{J}(\xi_0,\theta_0) = {\theta_0}_{XX} +{\theta_2}_{zz} -{\xi_0}_{X} T^b_{z}.
\end{aligned}
\end{equation}
Substituting the leading-order solution leads to 
\begin{equation}
\begin{aligned}
{\theta_2}_{zz}=-\Racp~(\overline{\theta_0}_{X})^2 \mathrm{P}^\prime - \overline{\theta_0}_{XX}\left[1-\Racp~P T^b_{z}\right].
\label{sol}
\end{aligned}
\end{equation}
By averaging \eqref{sol} over the height and applying the boundary conditions on $\theta$, we obtain
\begin{equation}
\int_{0}^{1}{\theta_2}_{zz}~\mathrm{d}z=[{\theta_2}_{z}]^{1}_{-1}=0,
\end{equation}
since $P(-1)=P(1)=0$. Seeking a non-zero leading order for the temperature field implies
\begin{equation}
\int_{0}^{1}\left[1-\Racp~PT^b_{z}\right]~\mathrm{d}z=0.
\label{crit}
\end{equation}
This leads to
\begin{equation}
\Racp = \frac{2880}{9-5f},
\label{Racp}
\end{equation}
which corresponds to the definition of $\Racp$ in \eqref{defRacp}. 
When $f=0$, $\Racp = 320$, corresponding to 16 times the value of 20 given in CP for mixed boundary conditions, also found by \cite{Soward_Oruba_Dormy_2022}.
In CP, the Rayleigh number is based on the half-height, so the height's fourth power in $\Ra$ results in a factor of 16.
Similarly, when $f=1$, we recover the threshold value of 720 found by \cite{Shivashinsky}.
It is noteworthy that the threshold increases as the normalized flux difference decreases.
For $f \neq 0$ fixed, this occurs because the heat flux decreases linearly with altitude, resulting in a fluid particle perceiving less buoyancy as it rises.
Consequently, the buoyancy force experienced by the particle diminishes, making it more likely to be stabilized by diffusive effects.

\begin{figure}
   \centering
    \includegraphics[scale=0.35]{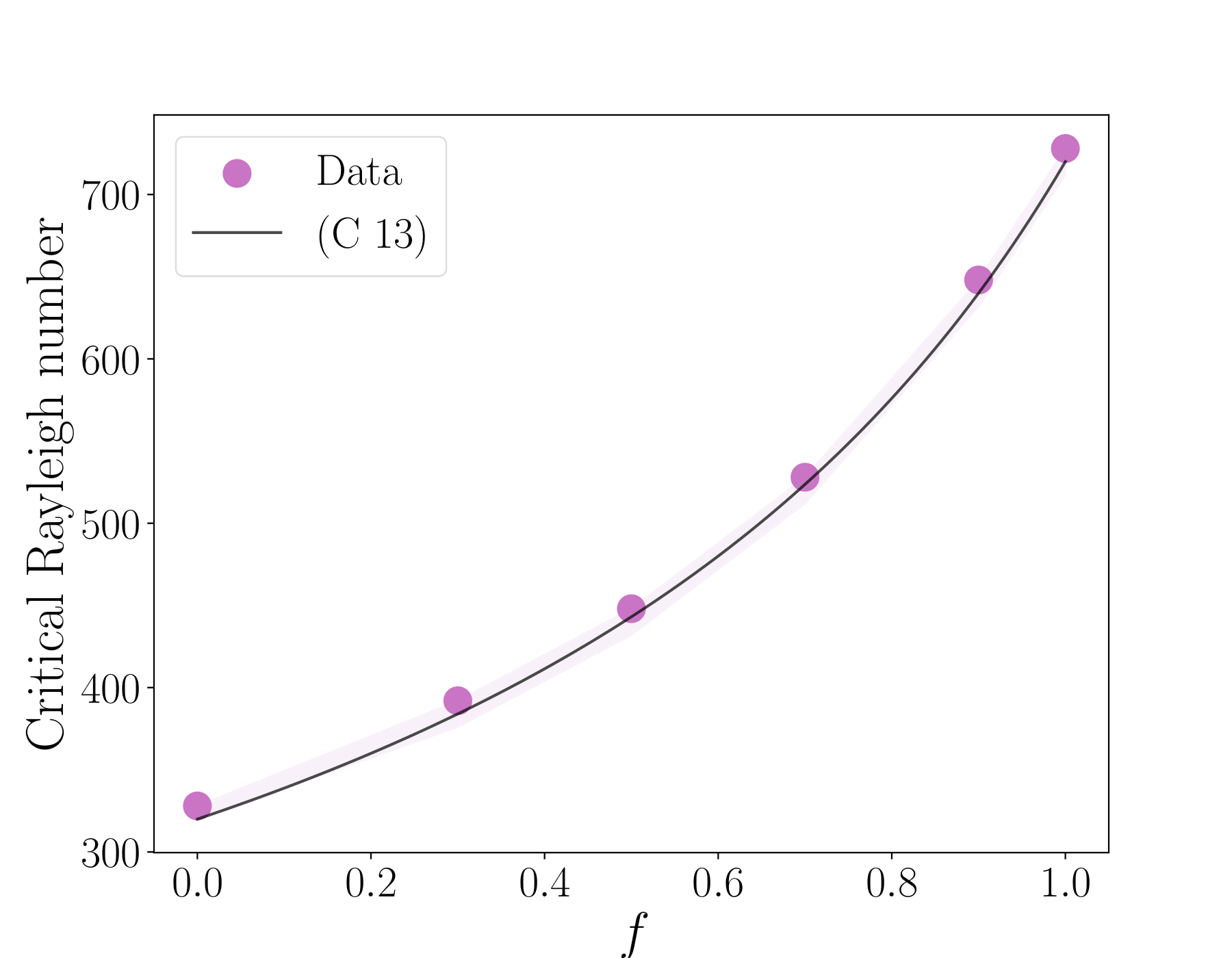}
   \caption{Critical Rayleigh number for the generalized Chapman \& Proctor instability as a function of the normalized flux difference.}
   \label{fig17}
\end{figure}

To verify the validity of \eqref{Racp}, we perform DNS simulations of the CP configuration, using the IVP-method, in a 2D box with an aspect ratio of 32 and periodic boundaries in the $x$-direction. 
We include a uniform volume heat sink term $f$ in the temperature equation and fix the heat fluxes to $1$ and $1-f$ to the bottom and the top respectively.
\autoref{fig17} shows the marginal stability curve i.e the $\Racp$ as a function of $f$.
Each data point was obtained by initiating a simulation from an initial condition with no flow and a zero temperature field in the domain, and then identifying the 
$\Ra$ value for which exponential growth in velocity magnitude first occurs. 
We observed no significant dependence on the aspect ratio.
The numerical data align perfectly with \eqref{Racp}.

\subsection{Flow stabilization effect}\label{StabCP}
\begin{figure}
   \centering
    \includegraphics[scale=0.2]{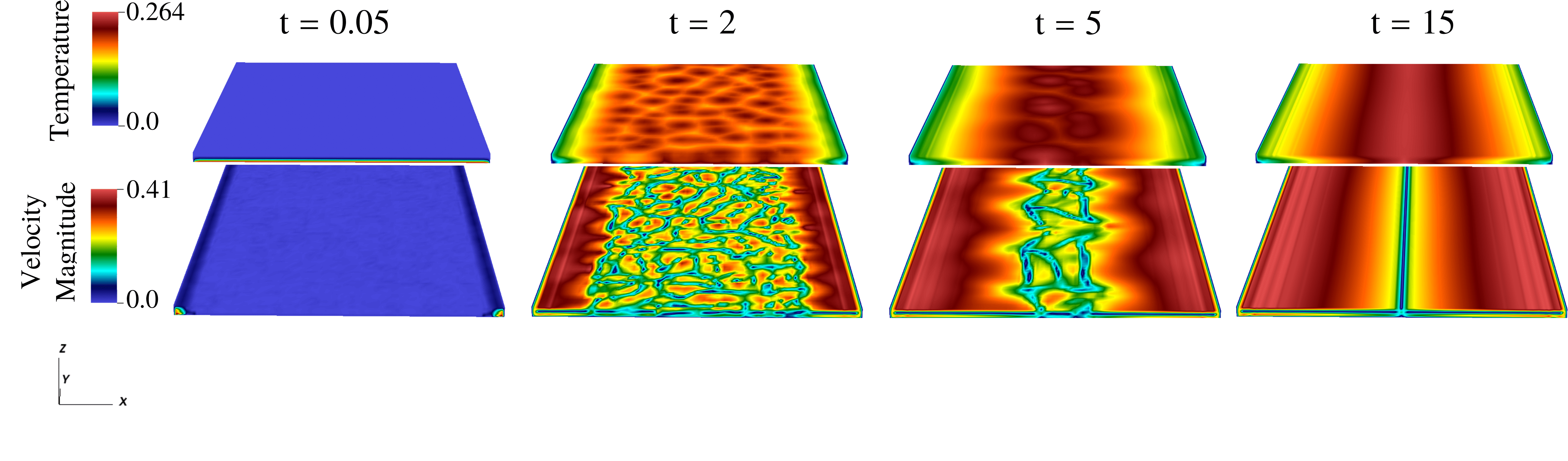}
   \caption{Snapshots of the 3D temperature (top) and velocity magnitude (bottom) fields, at various times.}
   \label{fig18}
\end{figure}

The generalized Chapman–Proctor instability can initially emerge in the system but may be suppressed during the transient phase of base flow development.
\autoref{fig18} presents snapshots from a 3D IVP showing the temperature field (top) and the velocity amplitude (bottom) at successive times.
The simulation parameters are $\epsilon = 1/16$, $\Ra = 5000$, and $f = 0.9$.
The Prandtl number is fixed at $Pr = 1$ to ensure that the base state remains stable with respect to the three-dimensional instability analyzed in \autoref{stabanalysis}.
We recall that IVP simulations are initiated with a quiescent fluid and a uniform temperature field $T=0$, where small temperature perturbations of amplitude $10^{-3}$ are introduced, leading to thermal convection growth during a transient phase, as shown in \autoref{fig18}.
We choose a sufficiently large $\Ra$ to be supercritical for the Chapman \& Proctor instability (threshold at $\Ra = 640 $ for $f=0.9$).
At the beginning, when the system is close to the initial conditions, 
the flow from the sidewall has not yet expanded throughout the entire domain and the temperature is mainly uniform. 
As time progresses, the Chapman \& Proctor instability emerges, forming convection rolls ($t=2$).
At this stage, the base flow has not yet propagated throughout the domain.
As this flow advances towards the centre, the convection rolls disappear, the system stabilizes ($t=5$) and becomes completely stable (we observe exponential decay of all fluctuations) when the steady state is reached ($t=15$).

\bibliographystyle{jfm}
\bibliography{jfm-instructions}
\end{document}